%% file: SUS-21-009_temp.tex
\pdfoutput=1
\documentclass[11pt,twoside,a4paper,cmspaper,final,collab]{cms-tdr}
\def\svnVersion{db4b2e1}\def\svnDate{2023/10/09}\def\cmsCernNoTag{CERN-EP-2023-127}\def\cmsCernDate{\today}\def\cmsMessage{Published in the Journal of High Energy Physics as \href{http://dx.doi.org/10.1007/JHEP10(2023)046}{\doi{10.1007/JHEP10(2023)046}.}}
\begin{document}\cmsNoteHeader{SUS-21-009}

\providecommand{\cmsTable}[1]{\resizebox{\textwidth}{!}{#1}}
\newcommand{\HsGra}{\ensuremath{\PH\sGra}\xspace}
\newcommand{\ZsGra}{\ensuremath{\PZ\sGra}\xspace}
\newcommand{\gsGra}{\ensuremath{\PGg\sGra}\xspace}
\newcommand{\ChiNOne}{\PSGczDo}
\newcommand{\ChiNTwo}{\PSGczDt}
\newcommand{\ChiCOne}{\PSGcpmDo}
\newcommand{\ChiCOneMP}{\ensuremath{\HepParticle{\PSGc}{1}{\mp}}\xspace}
\newcommand{\ChiCOneChiCOneMP}{\ensuremath{\ChiCOne\ChiCOneMP}\xspace}
\newcommand{\ChiCOneChiNOne}{\ensuremath{\ChiCOne\ChiNOne}\xspace}
\newcommand{\ChiCOneChiNTwo}{\ensuremath{\ChiCOne\ChiNTwo}\xspace}
\newcommand{\ChiNOneChiNTwo}{\ensuremath{\ChiNOne\ChiNTwo}\xspace}
\newcommand{\ChiNOneToHsGra}{\ensuremath{\ChiNOne\to\HsGra}\xspace}
\newcommand{\ChiNOneTogsGra}{\ensuremath{\ChiNOne\to\gsGra}\xspace}
\newcommand{\ChiNOneToZsGra}{\ensuremath{\ChiNOne\to\ZsGra}\xspace}
\newcommand{\ChiCOneToWsGra}{\ensuremath{\ChiCOne\to\PW\sGra}\xspace}
\newcommand{\GluinoToBBChiNOne}{\ensuremath{\PSg\to\bbbar\ChiNOne}\xspace}
\newcommand{\GluinoToTTChiNOne}{\ensuremath{\PSg\to\ttbar\ChiNOne}\xspace}
\newcommand{\GluinoToQQChiNOne}{\ensuremath{\PSg\to\qqbar\ChiNOne}\xspace}
\newcommand{\msGra}{\ensuremath{m_\sGra}\xspace}
\newcommand{\mJ}{\ensuremath{m_{\mathrm{J}}}\xspace}
\newcommand{\Pellprime}{\ensuremath{\Pell^{\prime}}\xspace}
\newcommand{\wlnudecay}{\ensuremath{\PW\to\Pellprime\PGn}\xspace}
\newcommand{\znunudecay}{\ensuremath{\PZ\to\PGn\PGn}\xspace}
\newcommand{\ttjets}{\ensuremath{\PQt\PAQt{+}\text{jets}}\xspace}
\newcommand{\zjet}{\ensuremath{\PZ{+}\text{jet}}\xspace}
\newcommand{\zjets}{\ensuremath{\PZ{+}\text{jets}}\xspace}
\newcommand{\wjets}{\ensuremath{\PW{+}\text{jets}}\xspace}
\newcommand{\ttgammaj}{\ensuremath{\PQt\PAQt\PGg{+}\text{jets}}\xspace}
\newcommand{\wgammaj}{\ensuremath{\PW\PGg{+}\text{jets}}\xspace}
\newcommand{\zgammaj}{\ensuremath{\PZ\PGg{+}\text{jets}}\xspace}
\newcommand{\ttgamma}{\ensuremath{\PQt\PAQt\PGg}\xspace}
\newcommand{\wgamma}{\ensuremath{\PW\PGg}\xspace}
\newcommand{\znunu}{\ensuremath{\PZ(\PGn\PGn)}\xspace}
\newcommand{\znunuj}{\ensuremath{\PZ(\PGn\PGn){+}\text{jets}}\xspace}
\newcommand{\znunugammaj}{\ensuremath{\PZ(\PGn\PGn)\PGg{+}\text{jets}}\xspace}
\newcommand{\zgamma}{\ensuremath{\PZ(\PGn\PGn)\PGg}\xspace}
\newcommand{\zllgamma}{\ensuremath{\PZ(\Pell\Pell)\PGg}\xspace}
\newcommand{\zeegammaj}{\ensuremath{\PZ(\Pe\Pe)\PGg{+}\text{jets}}\xspace}
\newcommand{\zmumugammaj}{\ensuremath{\PZ(\PGm\PGm)\PGg{+}\text{jets}}\xspace}
\newcommand{\zee}{\ensuremath{\PZ(\Pe\Pe)}\xspace}
\newcommand{\gjet}{\ensuremath{\PGg{+}\text{jet}}\xspace}
\newcommand{\gjets}{\ensuremath{\PGg{+}\text{jets}}\xspace}
\newcommand{\dphi}{\ensuremath{\Delta\phi}\xspace}
\newcommand{\dphifull}{\ensuremath{\dphi(\ptvec^{\text{jet}},\ptvecmiss)}\xspace}
\newcommand{\dR}{\ensuremath{\Delta R}\xspace}
\newcommand{\nj}{\ensuremath{N_{\text{jets}}}\xspace}
\newcommand{\nb}{\ensuremath{N_{\text{\PQb-tags}}}\xspace}
\newcommand{\ST}{\ensuremath{S_{\mathrm{T}}}\xspace}
\newcommand{\abseta}{\ensuremath{\abs{\eta}}\xspace}
\newcommand{\pp}{\ensuremath{\Pp\Pp}\xspace}

\newcommand{\cPV}{\ensuremath{\text{V}}\xspace}
\newcommand{\vtag}{\ensuremath{\cPV\text{-tag}}\xspace}
\newcommand{\htag}{\ensuremath{\PH\text{-tag}}\xspace}
\newcommand{\vtags}{\ensuremath{\cPV\text{-tags}}\xspace}
\newcommand{\htags}{\ensuremath{\PH\text{-tags}}\xspace}
\newcommand{\vorhtag}{\ensuremath{\cPV\text{- or }\PH\text{-tag}}\xspace}
\newcommand{\vandhtag}{\ensuremath{\cPV\text{- and }\PH\text{-tag}}\xspace}
\newcommand{\vandhtags}{\ensuremath{\cPV\text{- and }\PH\text{-tags}}\xspace}

\newcommand{\llg}{\ensuremath{\Pell\Pell\PGg}\xspace}
\newcommand{\lgamma}{\ensuremath{\Pell\PGg}\xspace}
\newcommand{\NdataSRlep}{\ensuremath{N^{\text{data}}_{0\lgamma}}\xspace}
\newcommand{\NdataCRlep}{\ensuremath{N^{\text{data}}_{1\lgamma}}\xspace}
\newcommand{\Qmult}{\ensuremath{Q_{\text{mult}}}\xspace}
\newcommand{\NdataSRefake}{\ensuremath{N^{\text{data}}_{\PGg}}\xspace}
\newcommand{\NdataCRefake}{\ensuremath{N^{\text{data}}_{1\Pe\,0\PGg}}\xspace}
\newcommand{\NdataCRzll}{\ensuremath{N^{\text{data}}_{\llg}}\xspace}
\newcommand{\NmcCRzll}{\ensuremath{N^{\text{MC}}_{\llg}}\xspace}
\newcommand{\NdataSRznunu}{\ensuremath{N^{\text{data}}_{\zgamma}}\xspace}
\newcommand{\NmcSRznunu}{\ensuremath{N^{\text{MC}}_{\zgamma}}\xspace}
\newcommand{\betazll}{\ensuremath{\beta_{\llg}}\xspace}
\newcommand{\Rlowmet}{\ensuremath{R_{\text{low-}\ptmiss}}\xspace}
\newcommand{\Rhighmet}{\ensuremath{R_{\text{high-}\ptmiss}}\xspace}
\newcommand{\muR}{\ensuremath{\mu_{\text{R}}}\xspace}
\newcommand{\muF}{\ensuremath{\mu_{\text{F}}}\xspace}
\newcommand{\DEEPCSV}{\textsc{DeepCSV}\xspace}
\newcommand{\muSig}{\ensuremath{\mu}\xspace}
\newcommand{\qmu}{\ensuremath{q_{\muSig}}\xspace}
\newcommand{\Lmax}{\ensuremath{\mathcal{L}_{\text{max}}}\xspace}
\newcommand{\Lmu}{\ensuremath{\mathcal{L}_{\mu}}\xspace}

\newcolumntype{A}{r@{\hspace{0em}}l}
\newcolumntype{B}{A@{$\,\pm\,$}A}

\hyphenation{electro-weak-ino}

\cmsNoteHeader{SUS-21-009}
\title{Search for new physics in multijet events with at least one photon and large missing transverse momentum in proton-proton collisions at 13\TeV}

\date{\today}

\abstract{
A search for new physics in final states consisting of at least one photon, multiple jets, and large missing transverse momentum is presented, using proton-proton collision events at a center-of-mass energy of 13\TeV. The data correspond to an integrated luminosity of 137\fbinv, recorded by the CMS experiment at the CERN LHC from 2016 to 2018. The events are divided into mutually exclusive bins characterized by the missing transverse momentum, the number of jets, the number of \PQb-tagged jets, and jets consistent with the presence of hadronically decaying \PW, \PZ, or Higgs bosons. The observed data are found to be consistent with the prediction from standard model processes. The results are interpreted in the context of simplified models of pair production of supersymmetric particles via strong and electroweak interactions. Depending on the details of the signal models, gluinos and squarks of masses up to 2.35 and 1.43\TeV, respectively, and electroweakinos of masses up to 1.23\TeV are excluded at 95\% confidence level.
}

\hypersetup{%
pdfauthor={CMS Collaboration},%
pdftitle={Search for new physics in multijet events with at least one photon and large missing transverse momentum in proton-proton collisions at 13 TeV},%
pdfsubject={CMS},%
pdfkeywords={CMS, SUSY, LHC}}

\maketitle

\section{Introduction} \label{sec:introduction}

Several extensions of the standard model (SM) of elementary particles attempt to provide an explanation for the origin of dark matter (DM)~\cite{Zwicky:1933gu,Rubin:1970zza} and to resolve the gauge hierarchy problem~\cite{Barbieri:1987fn,Dimopoulos:1995mi,Barbieri:2009ev,Papucci:2011wy}. Supersymmetry (SUSY)~\cite{Ramond:1971gb,Golfand:1971iw,Neveu:1971rx,Volkov:1972jx,Wess:1973kz,Wess:1974tw,Fayet:1974pd,Nilles:1983ge} is a hypothesized symmetry between fermions and bosons, which, when included in extensions to the SM, predicts a new bosonic (fermionic) superpartner for each SM fermion (boson). In SUSY models with $R$-parity conservation~\cite{bib-rparity}, the lightest supersymmetric particle (LSP) is stable and often neutral and weakly interacting, making it a possible DM candidate. In addition, gauge coupling unification is a natural possibility in supersymmetric theories~\cite{Emmanuel-Costa:2006aqu}.

Superpartners would contribute to quantum corrections to the Higgs boson (\PH) mass such that the \PH mass parameter would have only a logarithmic dependence on the scale of new physics. This effect could reduce the need for fine-tuning of the \PH mass~\cite{Papucci:2011wy}, thereby preserving naturalness, if the superpartners with the largest contributions to the corrections are sufficiently light. These particles include the gluino, top squark, and bottom squark, which are the superpartners of the SM gluon, top quark, and bottom quark, respectively.
If those color-charged superpartners are not accessible at the LHC, SUSY may still satisfy naturalness conditions if the higgsino, the superpartner of the Higgs boson, is near the electroweak scale~\cite{Baer:2012up}.

This paper explores the signatures of SUSY particles produced via strong and electroweak interactions in proton-proton (\pp) collisions, with a particular focus on final states containing at least one photon, multiple jets, and large missing transverse momentum. In gauge-mediated SUSY-breaking scenarios, the LSP is a gravitino (\sGra), the superpartner of the graviton, and it is expected to be roughly a few {\GeVns} in mass~\cite{Fayet:1977vd,Baer:1996hx}. If the next-to-LSP (NLSP) is a chargino (neutralino), its decay will result in a \PW boson (photon, \PZ boson, or Higgs boson) and a \sGra.
The neutralino is an admixture of neutral wino, bino, and/or higgsino components that can couple to photons.
Such decays are especially prominent if sleptons, the superpartners of SM leptons, are sufficiently massive that decays to sleptons are suppressed.

We interpret the results of this search using simplified models~\cite{bib-sms-1,bib-sms-2,bib-sms-3,bib-sms-4,Chatrchyan:2013sza} of
SUSY particle production via strong and electroweak interactions.
Specifically, we consider several models of squark- and gluino-mediated production of charginos and neutralinos, in which each of the latter particles subsequently decays to the LSP and an SM boson.
We also consider the production of neutralinos and charginos, collectively referred to as electroweakinos, via electroweak interactions.
In all simplified models considered in this paper, the decays of SUSY particles are assumed to be prompt, and the mass of the gravitino, \msGra, is fixed to be 1\GeV.
The event kinematic properties do not depend strongly on the exact choice of \msGra in the phase space explored in this analysis.
This search therefore targets final states with at least one photon produced from the decay of a neutralino and missing transverse momentum from the LSP, which escapes the collision region without detection.

\begin{figure}[h!]
  \centering
  \includegraphics[width=0.40\linewidth]{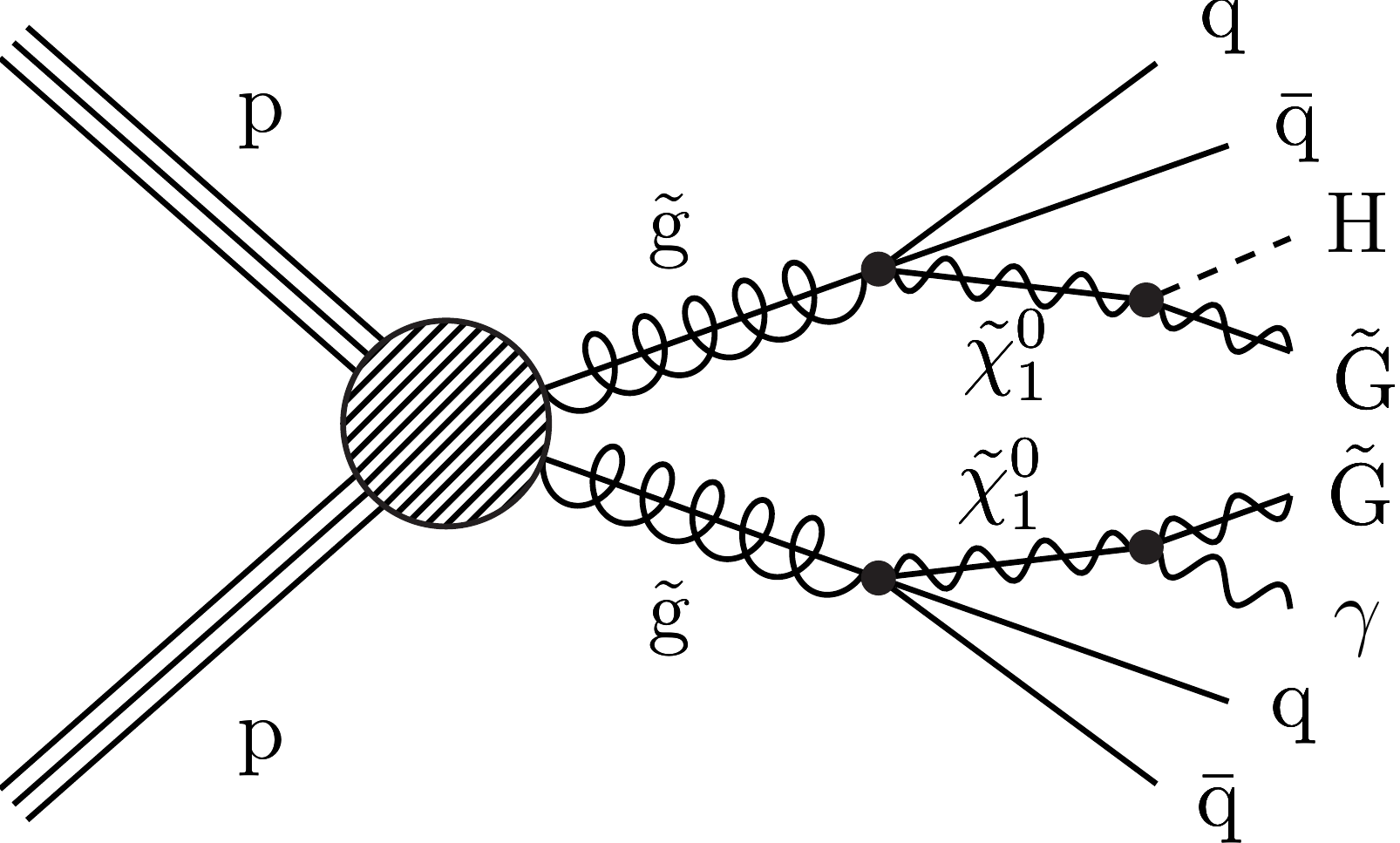}
  \includegraphics[width=0.40\linewidth]{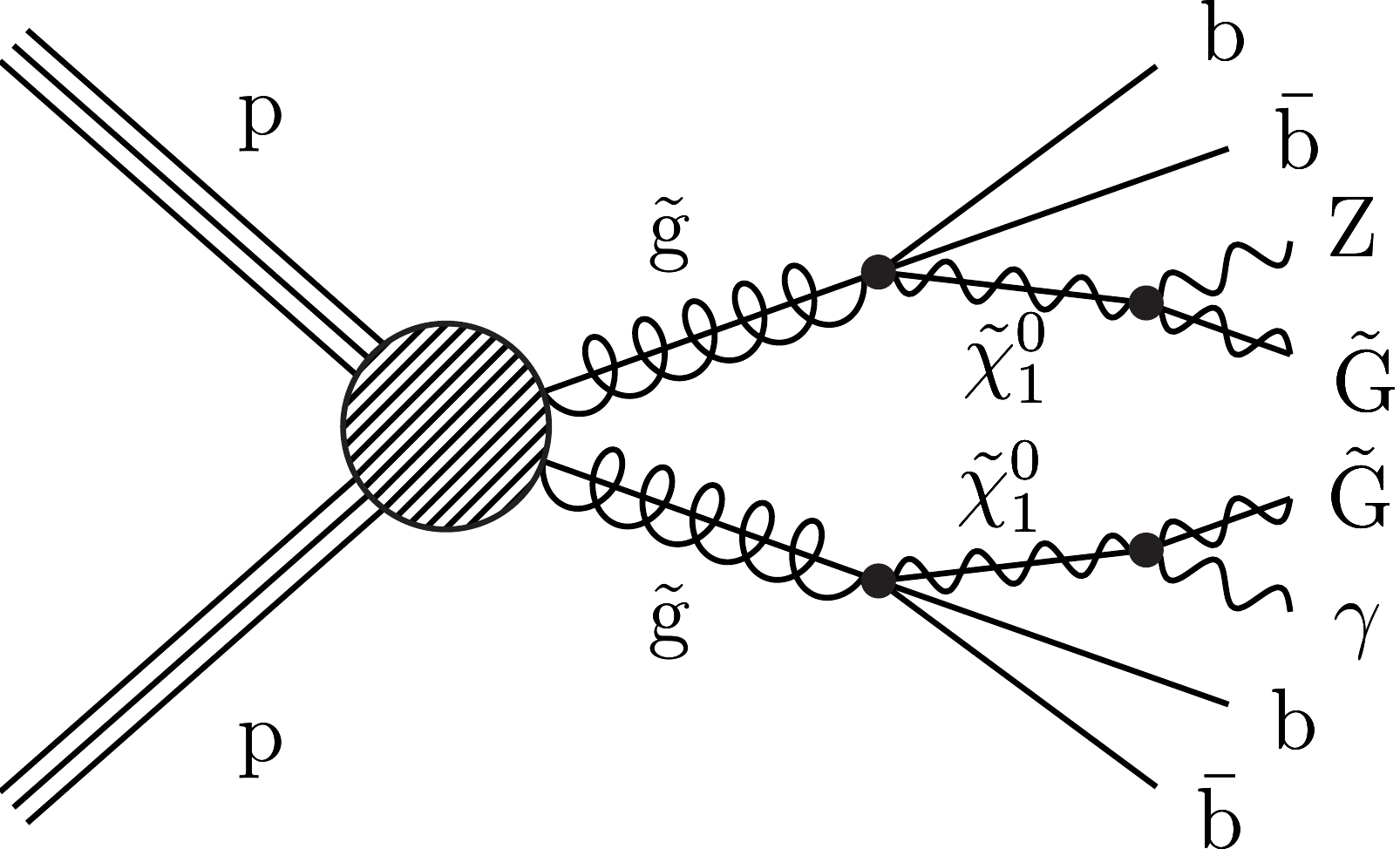}\\
  \includegraphics[width=0.40\linewidth]{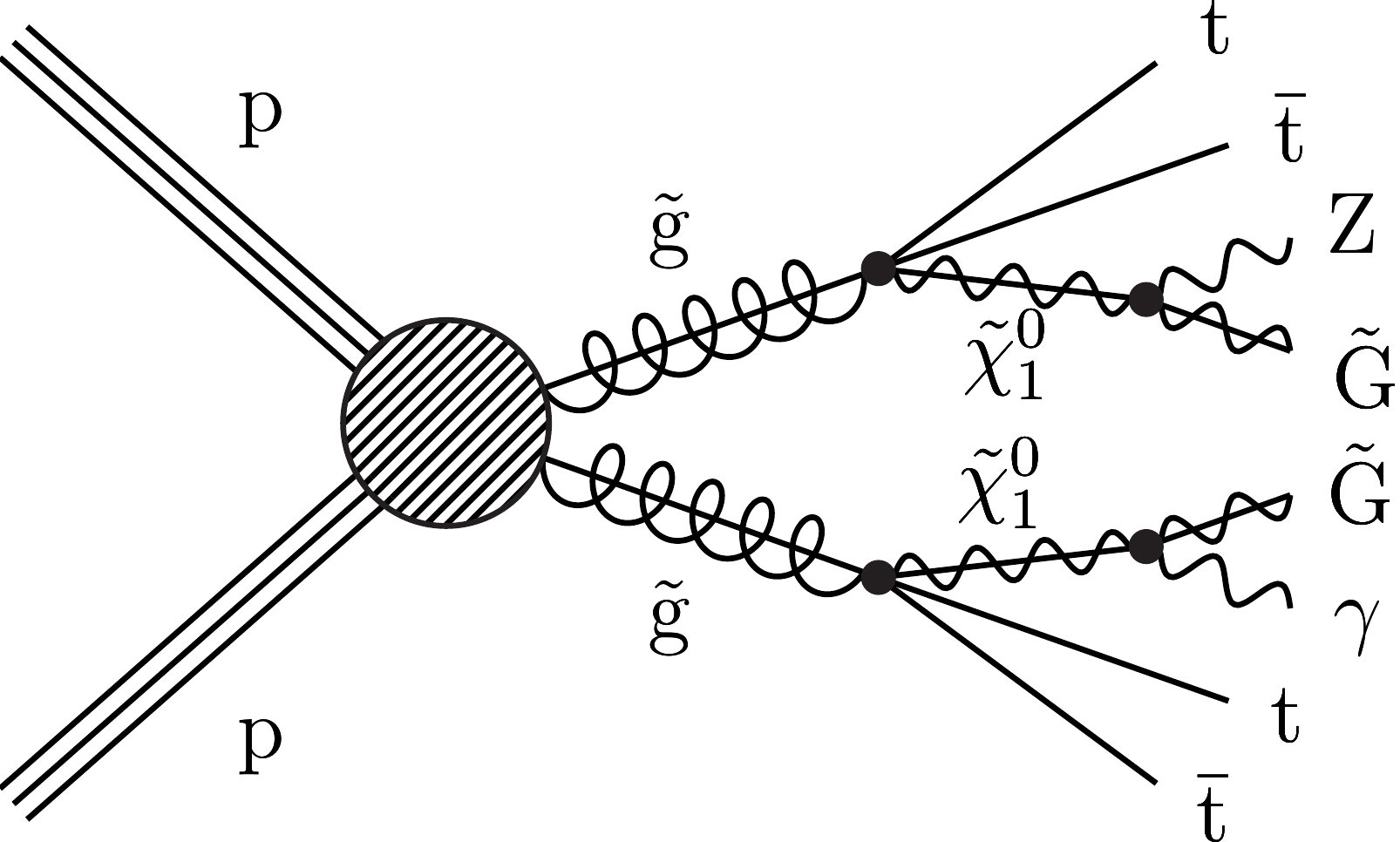}
  \includegraphics[width=0.40\linewidth]{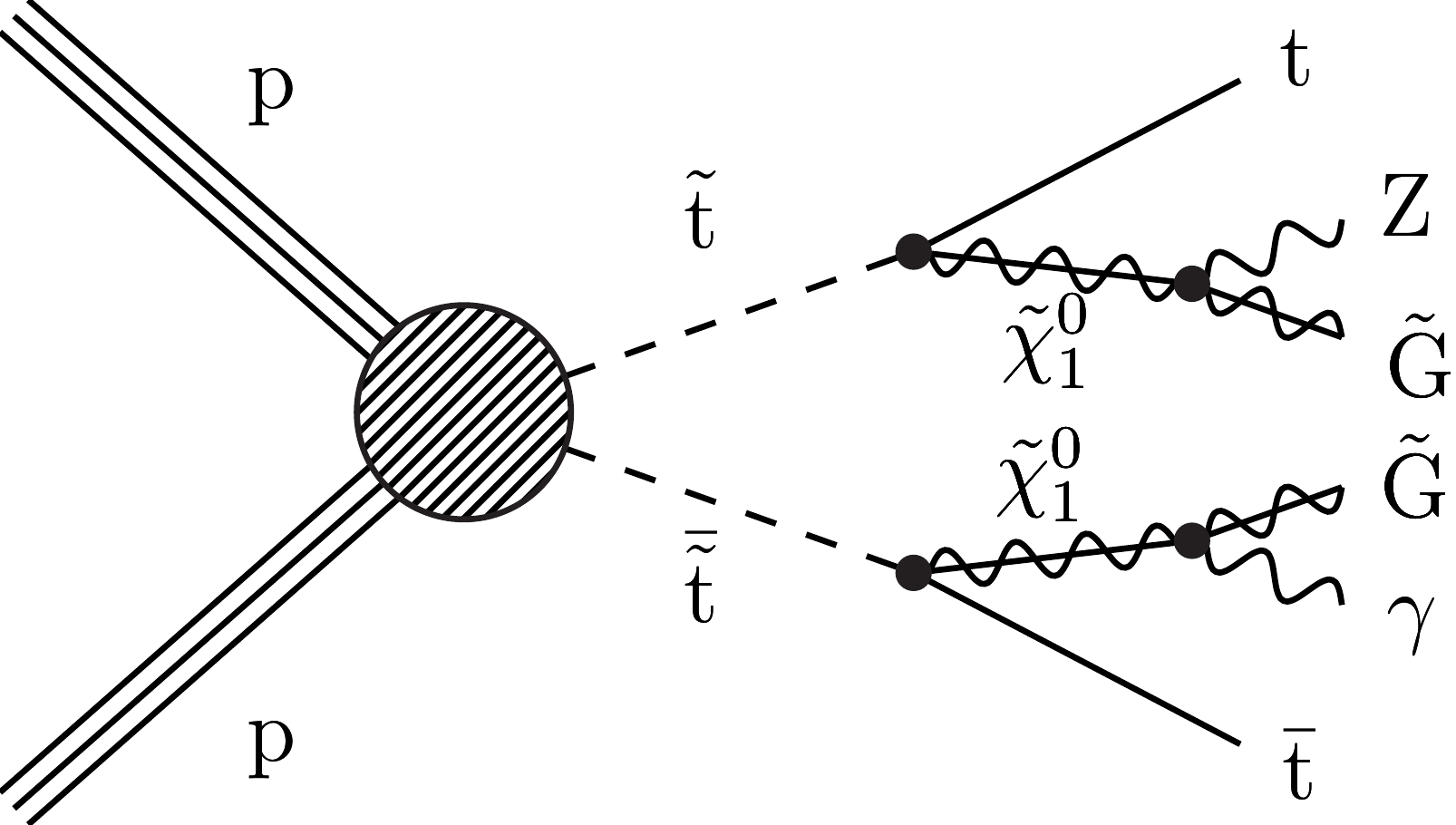}
  \caption{Diagrams of simplified models of gluino pair production: T5qqqqHG (upper left), T5bbbbZG (upper right), T5ttttZG (lower left), and top squark pair production: T6ttZG (lower right). The models are defined in the text.}
  \label{fig:strong-sms}
\end{figure}

Representative diagrams depicting simplified models~\cite{Chatrchyan:2013sza} of gluino (\PSg) pair production and top squark (\PSQt) pair production are shown in Fig.~\ref{fig:strong-sms}.
In the gluino pair production models, the gluino decays to a neutralino (\ChiNOne) and a pair of quarks; any possible \ChiNOne decays occur with equal probability.
The T5qqqqHG model is defined by the decay of gluinos to a pair of light-flavored quarks (\qqbar) and \ChiNOne, followed by \ChiNOneToHsGra or \ChiNOneTogsGra with 50\% branching fraction each.
The mass of the \ChiNOne is taken to be 127\GeV or above.
In the T5bbbbZG and T5ttttZG models, \GluinoToBBChiNOne and \GluinoToTTChiNOne, respectively, and the \ChiNOne decays to \ZsGra or \gsGra with 50\% branching fraction each.
In the top squark pair production model T6ttZG, each top squark decays to a top quark and a \ChiNOne, followed by a decay \ChiNOneToZsGra or \ChiNOneTogsGra with 50\% branching fraction each.
In the models involving \ChiNOneToZsGra, \ChiNOne masses as low as 10\GeV are considered.

\begin{figure}[h!]
  \centering
  \includegraphics[width=0.40\linewidth]{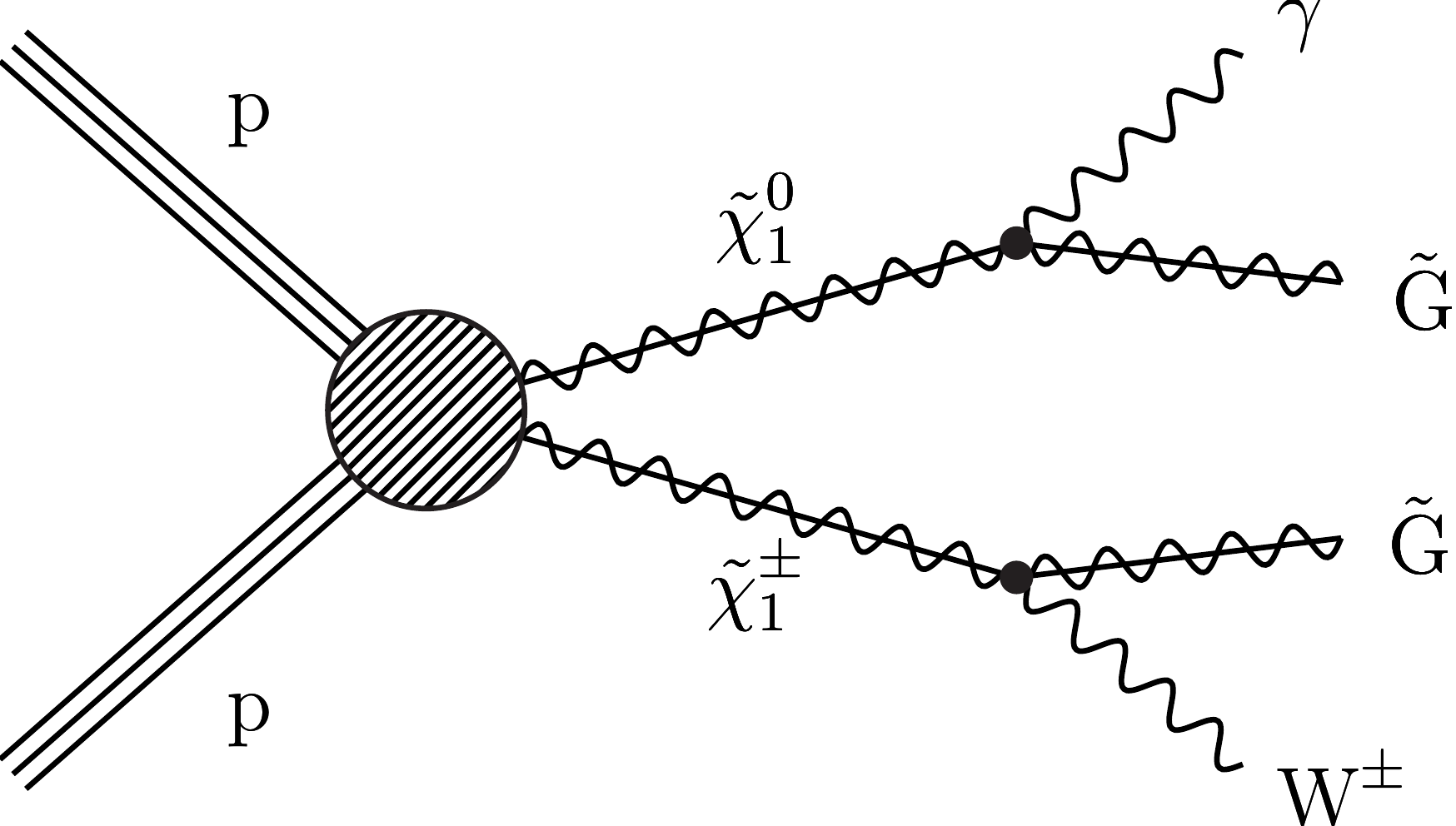}\\
  \includegraphics[width=0.40\linewidth]{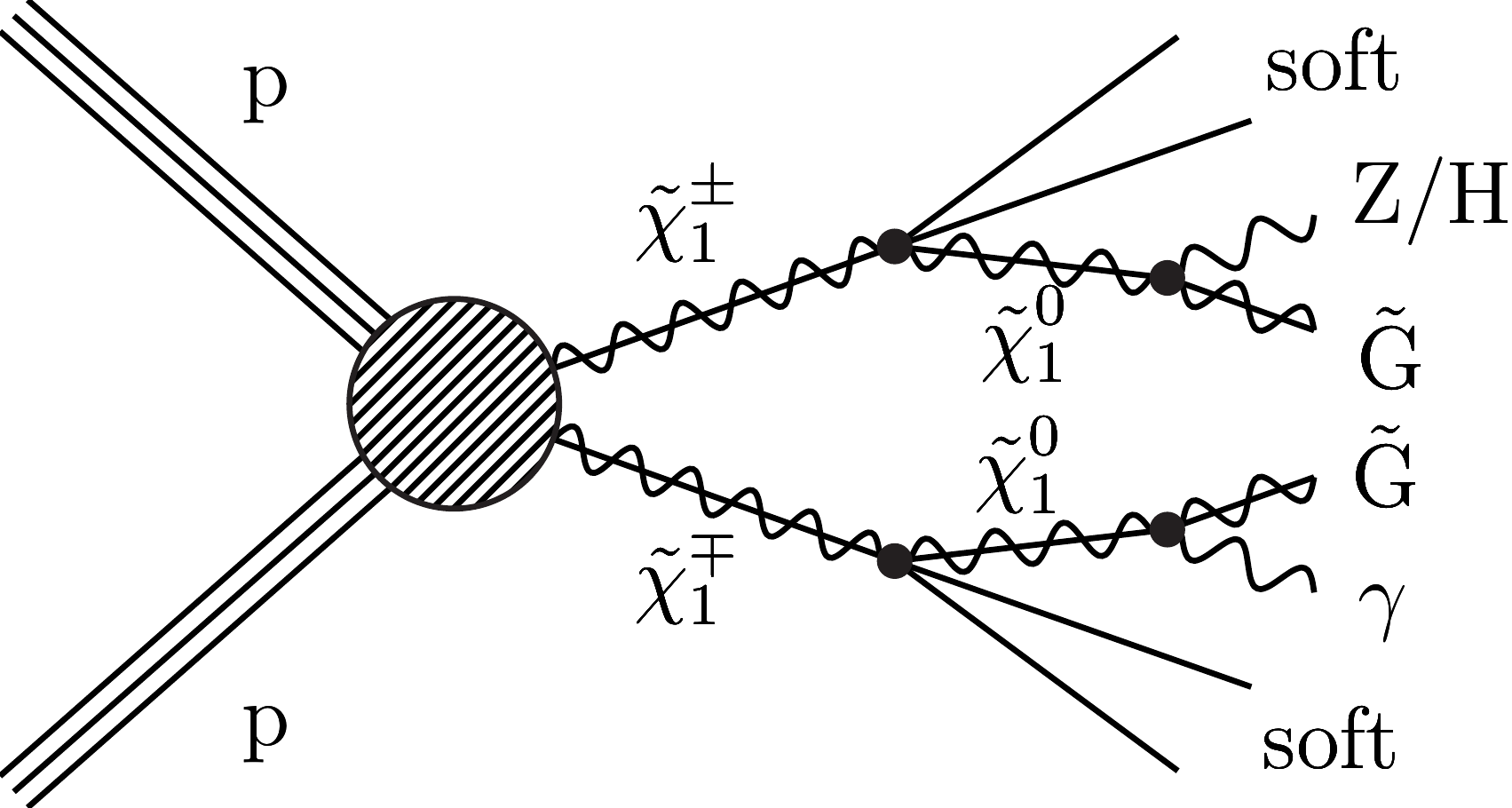}
  \includegraphics[width=0.40\linewidth]{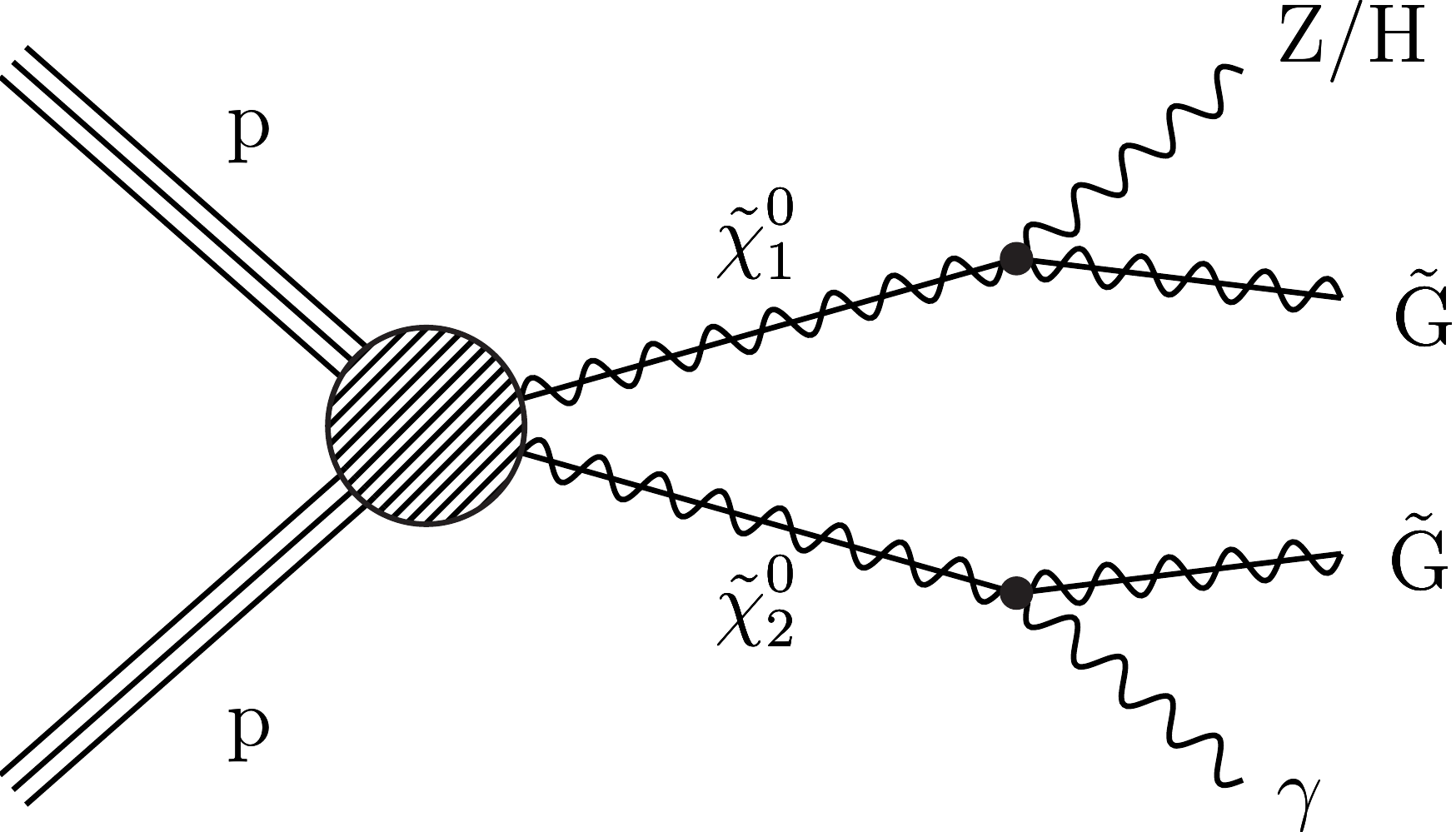}
  \caption{Diagrams of simplified models of electroweakino pair production: TChiWG (upper), TChiNG (lower left), and TChiNGnn (lower right). Only the \ChiCOneChiNOne and \ChiCOneChiCOneMP cases are shown for the TChiWG and TChiNG models, respectively. The models are defined in the text. ``Soft'' indicates particles with momentum too low to be detectable.}
  \label{fig:ewk-sms}
\end{figure}

Example diagrams for chargino-neutralino (\ChiCOneChiNOne) and chargino-chargino (\ChiCOneChiCOneMP) production are presented in Fig.~\ref{fig:ewk-sms},
and the models are denoted TChiWG and TChiNG, respectively. In both models, the charginos and neutralinos are degenerate in mass.
In the TChiWG model, the branching fractions for \ChiNOneTogsGra and \ChiCOneToWsGra are taken to be 100\%.
The TChiNG model includes all electroweak production modes of nearly degenerate triplet of chargino and neutralino states: \ChiCOneChiCOneMP, \ChiCOneChiNOne/\ChiCOneChiNTwo, and \ChiNOneChiNTwo.
The \ChiCOne and \ChiNTwo decay to \ChiNOne and low-momentum particles that are outside the kinematic acceptance of this analysis.
The \ChiNOne decays to \gsGra, \ZsGra, and \HsGra with branching fractions of 50, 25, and 25\%, respectively.
Additionally, there is a scenario denoted TChiNGnn, where \ChiNOne decays to \ZsGra and \HsGra with 50\% branching fraction each, and \ChiNTwo decays to \gsGra with 100\% branching fraction.
Only the \ChiNOneChiNTwo process contributes to this scenario.

Previous searches by the CMS Collaboration for signatures of squark, gluino, and electroweakino production involving photons
with data corresponding to an integrated luminosity of 36\fbinv at $\sqrt{s} = 13\TeV$ are documented in Refs.~\cite{Sirunyan:2017nyt,Sirunyan:2017yse,Sirunyan:2019vin}. Similar searches have also been reported by the ATLAS Collaboration based on data corresponding to integrated luminosities of 36~\cite{ATLASCollaboration:2016wlb,ATLASCollaboration:2018ggm} and 139\fbinv~\cite{ATLAS:2022ckd}. The present search follows the analysis strategy used in Ref.~\cite{Sirunyan:2019vin}. The search regions are reoptimized based on the availability of four times more data and are extended to include jets consistent with hadronic decays of \PW, \PZ, or Higgs bosons for electroweakino searches.

This paper is structured as follows. A brief description of the CMS detector and event reconstruction is given in Section~\ref{sec:det-reco}. The data sets are described in Section~\ref{sec:dataset} and the event selection in Section~\ref{sec:evtsel}. The methods used to estimate SM backgrounds are presented in Section~\ref{sec:smbkg} and the systematic uncertainties in the predictions are provided in Section~\ref{sec:syst}. The results and summary are presented in Sections~\ref{sec:results} and~\ref{sec:summary}, respectively.
Tabulated results are provided in the HEPData record for this search~\cite{hepdata}.

\section{The CMS detector and event reconstruction} \label{sec:det-reco}

The central feature of the CMS apparatus is a superconducting solenoid of 6\unit{m} internal diameter, providing a magnetic field of 3.8\unit{T}.
Within the solenoid volume are a silicon pixel and strip tracker, a lead tungstate crystal electromagnetic calorimeter (ECAL), and a brass and scintillator hadron calorimeter, each composed of a barrel and two endcap sections.
The tracker systems cover a pseudorapidity range of $\abseta < 2.5$, and the calorimeter systems cover $\abseta < 3.0$.
Forward calorimeters extend the coverage provided by the barrel and endcap detectors up to $\abseta < 5.2$.
Muons are measured in gas-ionization detectors embedded in the steel flux-return yoke outside the solenoid. A more detailed description of the CMS detector, together with a definition of the coordinate system used and the relevant kinematic variables, can be found in Ref.~\cite{CMS:2008cmsdet}. 
Events of interest are selected using a two-stage trigger system, described in Ref.~\cite{CMS:2016trigHLT}.

Collision events are reconstructed using the particle-flow (PF) algorithm~\cite{CMS:2017pflow} which combines information from various subdetectors in an optimized way to give a list of PF candidates, namely photons, electrons, muons, charged hadrons, and neutral hadrons. Charged-particle tracks are used to reconstruct \pp interaction vertices in the event. The primary vertex is taken to be the vertex corresponding to the hardest scattering in the event, evaluated using tracking information alone, as described in Section 9.4.1 of Ref.~\cite{CMS-TDR-15-02}. This vertex is required to be within 24\cm of the center of the detector in the $z$ direction, and within 2\cm in the transverse direction. The remaining reconstructed vertices are referred to as pileup vertices and correspond to additional \pp interactions in the same bunch crossing. 

Reconstructed PF candidates are clustered into jets using the infrared and collinear safe anti-\kt algorithm~\cite{Cacciari:2008gp, Cacciari:2011ma} with a distance parameter of 0.4 (0.8), referred to as AK4 (AK8) jets. The pileup contribution to the AK4 jet momentum is mitigated by discarding all the charged-particle tracks associated with pileup vertices and applying an offset correction to mitigate the average contribution of neutral particles~\cite{Cacciari:2007fd,CMS:2020ebo}. To mitigate the effect of pileup interactions on the AK8 jet momentum, a pileup per particle identification algorithm~\cite{CMS:2020ebo} is used, which makes use of local shape information to distinguish particles originating from hard scatter and pileup interactions. To account for the nonuniformity of the detector response across the jet \pt and $\eta$ ranges, jet energy corrections (JECs) are derived from the simulation in order to make the response of reconstructed jets equal to the particle-level jets on average. Residual differences in response between data and simulation are corrected based on dedicated measurements of the momentum balance in dijet, \gjet, \zjet, and multijet events~\cite{CMS:2016jec}. The jet energy resolution (JER) in the simulation is also modified by smearing the jet \pt, using scale factors derived from data ranging from 1.1--1.2. Jets potentially dominated by contributions from anomalous detector signals or reconstruction failures are discarded using dedicated jet identification (ID) criteria~\cite{CMS-PAS-JME-16-003}.
The AK4 jets used in this search are required to have $\pt > 30\GeV$ and $\abseta < 2.4$.

The hadronization products from a hadronic decay of an energetic \PW, \PZ, or Higgs boson can be clustered into a single wide jet, which is reconstructed as an AK8 jet.
The AK8 jets considered in this search are required to have $\pt > 200\GeV$ and $\abseta < 2.4$.
The AK8 jet mass, \mJ, is reconstructed using the soft-drop algorithm~\cite{Larkoski:2014wba}, which improves jet mass resolution by removing soft and wide-angle contributions.
Requirements are applied to \mJ to identify \PW, \PZ, or Higgs boson candidates.
The same requirement $65 < \mJ < 105\GeV$ is used for both \PW and \PZ bosons, as the difference in the particle masses is smaller than the \mJ resolution;
jets that pass this requirement are called \vtags, where \cPV = \PW or \PZ.
For Higgs bosons, the requirement $105 < \mJ < 140\GeV$ is applied, and passing jets are called \htags.

Jets originating from {\PQb} quarks are identified by a combined secondary vertex algorithm based on a deep neural network (\DEEPCSV), applied to the reconstructed AK4 jets~\cite{Sirunyan:2017ezt}. A medium working point is used for the \DEEPCSV discriminator. This corresponds to a {\PQb} jet tagging efficiency of 65\% for jets with $\pt > 30\GeV$ with corresponding misidentification probability for gluon and light-quark (charm-quark) jets of 1.6\% (13\%).

The negative vector \ptvec sum of all PF particles is defined as the \ptvecmiss, and its magnitude is the missing transverse momentum (\ptmiss) used in this analysis. The \ptvecmiss is corrected for the changes in the \ptvec of jets after applying JECs. Events in which \ptmiss is identified to be originating from a mismeasured jet, detector noise, nonfunctional calorimetric channels, or reconstruction failures are rejected by dedicated algorithms~\cite{CMS:2019met}.
Events with $\ptmiss > 200\GeV$ are used in this analysis.

In order to improve the quality of the PF reconstruction, additional identification criteria are applied to photon candidates~\cite{CMS:2020egm}.
To suppress the misidentification of neutral pions, which are copiously produced in jets, as photons, the reconstructed photon candidates are required to be isolated. The isolation variable is defined as the \pt sum of a given type of PF particle candidate within a cone of radius $\dR=0.3$, where $\dR = \sqrt{\smash[b]{(\Delta\eta)^2+(\dphi)^2}}$ and $\phi$ is the azimuthal angle in radians, around the direction of the photon, excluding the photon itself. Separate isolation sums are computed for charged hadrons, photons, and neutral hadrons. The isolation sums are adjusted using $\eta$-dependent effective-area corrections, which account for the variation of energy density in a given event based on pileup interactions~\cite{CMS:2020egm}. The charged-hadron isolation sum is required to be less than 1.694 (2.089)\GeV for photons in the ECAL barrel (endcap); the requirements for the other isolation sums are expressed as linear or quadratic functions of the photon \pt for photon or neutral-hadron isolation, respectively. Isolated photons with $\pt > 100\GeV$ and $\abseta < 2.4$, excluding the ECAL barrel-endcap transition region $1.44 < \abseta < 1.56$, are considered for further analysis. The efficiency of the photon identification and isolation requirements is 90\% and the misidentification rate is 15--25\%.

As with photons, additional identification criteria are applied to electron~\cite{CMS:2020egm} and muon~\cite{Sirunyan:2018fpa} candidates.
The electron and muon candidates are also required to have $\pt>10\GeV$ and to be isolated, based on a variable defined as the scalar \pt sum of the charged hadron, neutral hadron, and photon PF candidates within a variable-radius cone around the lepton direction, divided by the lepton \pt. The expected contributions of neutral particles from pileup interactions are subtracted from the isolation sum~\cite{Cacciari:2007fd}. The radius of the cone is 0.2 for lepton $\pt<50\GeV$, $10\GeV/\pt$ for $50<\pt<200\GeV$, and 0.05 for $\pt>200\GeV$. The decreasing radius of the cone with lepton \pt is motivated by the increased collimation of the decay products from the lepton's parent particle with increasing Lorentz boost and helps to retain high lepton isolation efficiency in events with large number of jets or pileup interactions. The isolation variable is required to be less than 0.1 (0.2) for electrons (muons).

Charged-particle tracks, subsequently referred to as tracks, are used to reject events potentially containing hadronic decays of $\tau$ leptons; electrons or muons that could not be identified with the criteria described earlier; and low-momentum leptons from hadron decays.
An isolation requirement is applied to these tracks, based on the \pt sum of other tracks within a cone of 0.3 around each track.
The isolation sum divided by the track \pt is required to be less than 0.2 (0.1) for tracks identified as a PF electron or muon (PF charged hadron). The leptonic tracks are required to have $\pt > 5 \GeV$ and the hadronic tracks $\pt > 10\GeV$, along with the transverse mass between the track \pt and \ptmiss, defined as $\mT = \sqrt{2\pt^{\Pell} \ptmiss (1-\cos\dphi)}$, less than 100\GeV. Here, \dphi is the angular separation between the track \ptvec and the \ptvecmiss.
This \mT requirement preferentially selects isolated tracks from $\PW\to\Pell\PGn$ decays.

\section{Collision and simulated data sets}
\label{sec:dataset}

We select collision events that are recorded based on the \ptmiss reconstructed at the trigger level.
The trigger-level \ptmiss threshold varies from 90--140\GeV for the data set used in this search. 
The efficiencies of these \ptmiss triggers are measured using data sets recorded with single-electron triggers, as a function of the reconstructed \ptmiss.
Trigger efficiencies for events with reconstructed \ptmiss of at least 200 (300)\GeV are found to be 70, 60, and 60\% (95, 95, and 97\%) for data collected in 2016, 2017, and 2018, respectively. 
In addition to the \ptmiss-triggered data set, single-electron and single-muon data sets are also used for the estimation of the background from electrons misidentified as photons, described in Section~\ref{subsec:efake}. 
The estimation of backgrounds primarily relies on observed data, with Monte Carlo (MC) events used to derive correction factors and scale factors, to validate various background estimation methods, and to assess systematic uncertainties.

We use the \MGvATNLO event generator (version 2.2.2 for 2016, and 2.4.2 for 2017--2018)~\cite{Alwall:2014hca,Alwall:2007fs} at leading order (LO) precision for simulating the production of the \ttjets, \wjets, \zjets, QCD multijet, and \wgamma processes and at next-to-LO (NLO) precision for the \ttgamma process. The \ttgamma events are simulated with up to one additional parton at the matrix element level, while the \wgamma events have up to two additional partons, the \ttjets events up to three additional partons, and the other samples up to four additional partons. The single top quark process is modeled at NLO in perturbative QCD, with \MGvATNLO used for $s$-channel production. Signal events are generated with the \MGvATNLO generator at LO precision in a manner similar to the SM backgrounds, with up to two additional partons at the matrix element level. The decays of gluinos, top squarks, and neutralinos are modeled with {\PYTHIA}8~\cite{Sjostrand:2014zea}.

The NNPDF3.0 LO (NLO) parton distribution functions (PDFs) are used for samples simulated at LO (NLO) precision that correspond to the 2016 data~\cite{NNPDF:2014otw}. The NNPDF3.1 next-to-NLO (LO) PDFs are used for all 2017--2018 simulated background (signal) samples~\cite{Ball:2017nwa}. Parton showering and hadronization are performed for background samples using the \PYTHIA 8.212 generator~\cite{Sjostrand:2014zea} with the CUETP8M1 underlying event tune~\cite{Khachatryan:2015pea} for 2016 and \PYTHIA 8.226 (8.230) with the CP5 underlying event tune for 2017 (2018)~\cite{CMS:2019csb}. For the signal samples, the CUETP8M1 (CP2) underlying event tune and \PYTHIA version 8.226 (8.230) are used for 2016 (2017--2018). Partons generated with \MGvATNLO and \PYTHIA that would otherwise be counted twice are removed using the MLM~\cite{Alwall:2007fs} and FxFx~\cite{bib-merge} matching schemes in LO and NLO samples, respectively. The cross sections used for normalizing the signal yields are computed at NLO plus next-to-leading logarithmic (NLL) precision~\cite{bib-nlo-nll-03,bib-nlo-nll-04,Beenakker:2011sf,Beenakker:2013mva,Beenakker:2014sma,Beenakker:2016gmf,Beenakker:2016lwe}. 

The SM MC events are processed through a detailed simulation of the CMS detector based on the \GEANTfour~\cite{Agostinelli:2002hh} software. The simulated events are then reconstructed using the same algorithms as used for the collision data. 
The detector simulation of signal events is performed with the CMS fast simulation package~\cite{Abdullin:2011zz,Giammanco:2014bza}. 
The signal samples are corrected for differences with respect to the \GEANTfour-based simulation. 
Both the SM background samples and the signal samples are generated with nominal distributions of the number of pileup interactions, which are then reweighted to match the distribution measured in data.

\section{Event selection}
\label{sec:evtsel}

Events are selected using a set of criteria, referred to as the baseline selection, summarized in Table~\ref{tab:sel}. The events are required to have $\ptmiss > 300\GeV$, and $\nj\geq 2$, where \nj is the number of AK4 jets with $\pt>30\GeV$ and $\abseta<2.4$. The events must also contain at least one photon with $\pt > 100\GeV$ and $\abseta<2.4$. The photons are required to be separated from the jets by $\dR(j,\PGg)>0.3$.
The scalar \pt sum of the jets and the photon, denoted \ST, reflects the visible energy scale of the event and is required to be larger than 300\GeV.
Significant mismeasurement of jet \pt can lead to high \ptmiss, which is generally aligned with the mismeasured jet.
To suppress such events, we require $\dphifull>0.3$ for the two highest \pt jets in the event.
As the final state targeted contains only photons and hadronic jets, events with isolated electron and muon candidates are rejected to suppress the SM background arising from leptonic \PW decays.
Further, events with an isolated leptonic or hadronic track are rejected, eliminating an additional ${\approx}40\%$ of the relevant SM background processes.
The event samples used for estimation of backgrounds are referred to as control regions (CRs) and are defined to be nonoverlapping with the baseline selection.
These are designed---based on the presence of leptons (\Pe or \PGm), lower \ptmiss values, or lower \dphifull values---to be dominated by SM processes and are expected to have small signal contributions.

\begin{table}[h!]
\centering
\topcaption{Summary of the baseline selection criteria used to identify events of interest for this search.}
\label{tab:sel}
\begin{tabular}{ll} 
  \hline
  \ptmiss [{\GeVns}] &   $>300\GeV$ \\
  \nj ($\pt > 30\GeV$, $\abs{\eta} < 2.4$)              &   $\geq2$ \\
  \PGg ($\pt > 100\GeV$, $\abs{\eta} < 2.4$)         &   $\geq1$ \\
  $\ST = \sum_{\text{jets}}\pt + \pt^{\PGg}$ [{\GeVns}] &   $>300\GeV$ \\
  $\dphifull$  & $>0.3$ for two highest \pt jets \\
  Number of leptons (\Pe, $\mu$)     & $=0$ \\
  Number of isolated tracks   &  $=0$ \\
 \hline
\end{tabular}
\end{table}

The events satisfying the baseline criteria are further classified into mutually exclusive signal regions (SRs) to enhance the sensitivity of the analysis to different signal scenarios.
We define two sets of SRs called the electroweak (EW) and strong production (SP) SRs to target different types of signal models.
The EW SRs include events with $2 \leq \nj \leq 6$ and at least one \vorhtag.
These SRs are sensitive to electroweakino models like TChiWG, in which an energetic \PW, \PZ, or Higgs boson is expected in addition to low \nj.
They are also sensitive to gluino production models like T5qqqqZg in scenarios in which the difference between the masses of the gluino and the NLSP is small, resulting in soft jets and a boosted massive vector boson. 
The SP SRs include all baseline events not satisfying the EW SR selection criteria and are sensitive to gluino and squark pair production.

\begin{figure}[h]
\includegraphics[width=0.49\textwidth]{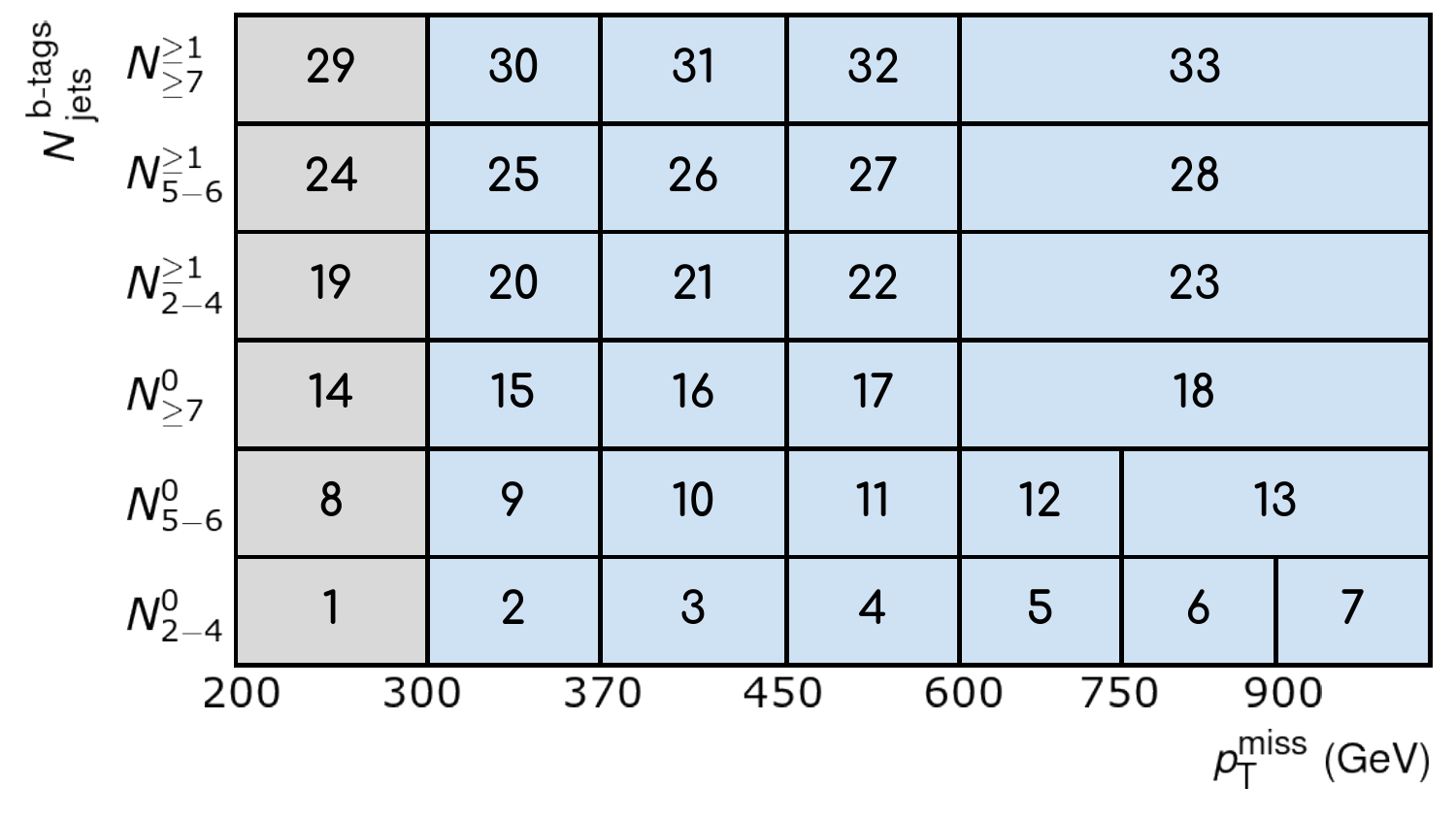}
\includegraphics[width=0.49\textwidth]{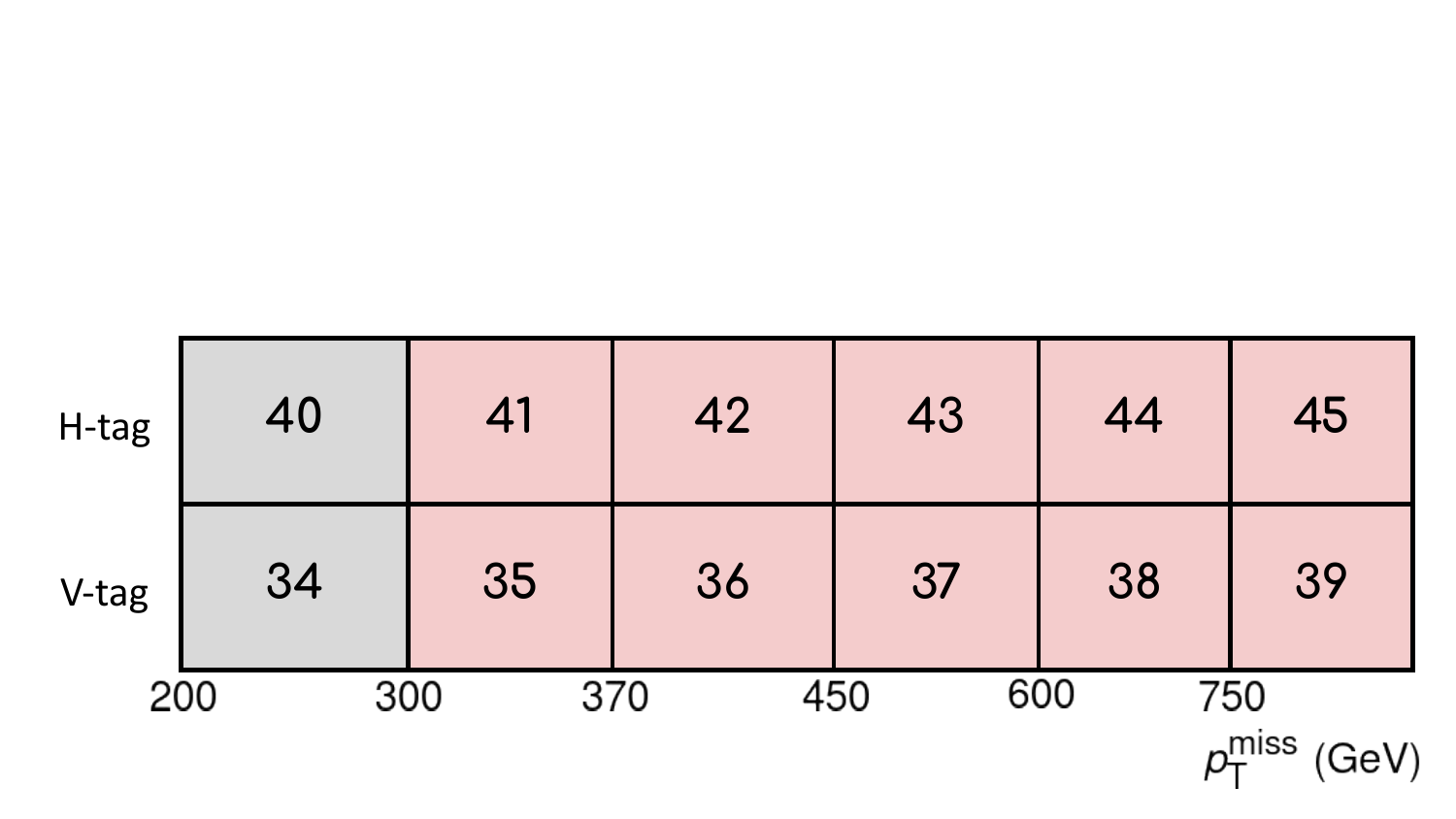}
\caption{The definitions and indexing schemes for the SP (left) and EW (right) SRs and low-\ptmiss CRs, in the planes of \ptmiss, \nj, and \nb (left) and \ptmiss, \vandhtag (right).
The gray blocks correspond to the low-\ptmiss CRs, the blue blocks to the SP SRs, and the red blocks to the EW SRs.}
\label{fig:Binning}
\end{figure}

Both the SP and EW SRs are binned in \ptmiss, with lower bin edges of 300, 370, 450, 600, 750, and 900\GeV.
These bins are chosen to ensure that each has an appreciable number of expected SM background events.
For the SP SRs, each of the \ptmiss bins is further divided into bins of \nj = 2--4, 5--6, and $\geq$ 7, and \nb = 0 and $\geq$ 1, where \nb is defined as the number of \PQb-tagged jets in the event.
For events with \nb = 0 and \nj = 5--6, the two highest \ptmiss bins are combined.
For events with \nb = 0 and \nj $\geq$ 7, and for all events with \nb $\geq$ 1, the three highest \ptmiss bins are combined.
The EW \ptmiss bins are defined based on the presence of a \vorhtag.
This scheme results in a total of 27 SP and 10 EW SRs, all statistically independent.
In addition, eight low-\ptmiss CRs are defined with $200 < \ptmiss < 300\GeV$ in the aforementioned bins of \nj and \nb, or \vandhtags, to be used for estimating the QCD multijet background.
The definitions of all the SP and EW SRs and the low-\ptmiss CRs are summarized in Fig.~\ref{fig:Binning}.
The indexing scheme shown in Fig.~\ref{fig:Binning}, with bin indices ranging from 1--45, is used to identify the SRs and low-\ptmiss CRs in the results presented in the following sections.
The other CRs are explained in more detail in Section~\ref{sec:smbkg}.

\section{Background estimation}
\label{sec:smbkg}

There are several SM processes that can result in final states containing at least one high-\pt photon, \ptmiss, and multiple jets. The production of \wgammaj, \ttgammaj, \wjets, \ttjets, \zgammaj, \gjets, and QCD multijet events are all non-negligible backgrounds for this search. 
The \wgammaj and \ttgammaj events, with a prompt photon and a \wlnudecay (\Pellprime = \Pe, \PGm, and \PGt leptons) decay, enter the search regions as the ``lost-lepton background'' if the \Pe or \PGm leptons are not identified, and therefore cannot be vetoed, or the \PGt leptons decay hadronically (\tauh).
The \wjets and \ttjets events contribute to the background if an electron originating from a \PW boson decay is misidentified as a photon. 
The \zgammaj process, with \znunudecay, is an irreducible background to this search. 
In all these processes, the presence of one or more neutrinos in the final states is the main source of \ptmiss.
In \gjets events, a \pt mismeasurement of one or more jets can lead to large \ptmiss in the reconstructed events.
The QCD multijet background arises similarly from artificial \ptmiss because of mismeasurement, but also requires that a jet be misidentified as a photon.

\subsection{Lost-lepton background}
\label{subsec:lostlepton}
\begin{figure*}[h!]
  \centering
  \includegraphics[width=0.49\linewidth]{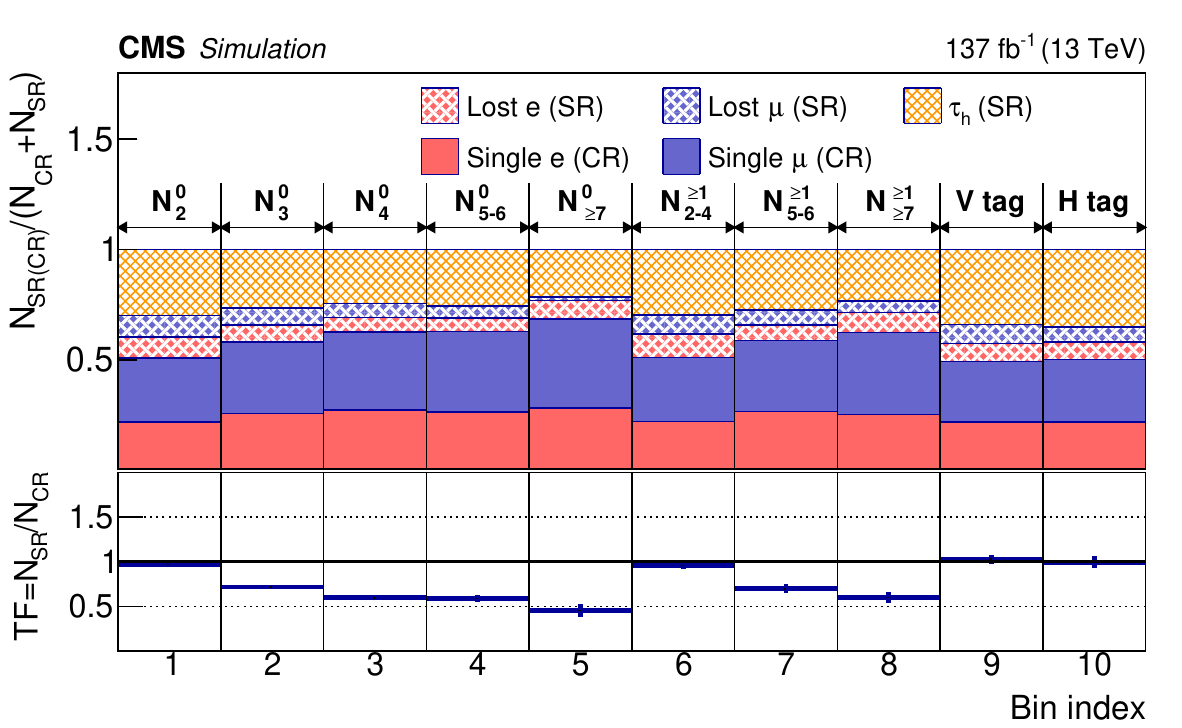}
  \includegraphics[width=0.49\linewidth]{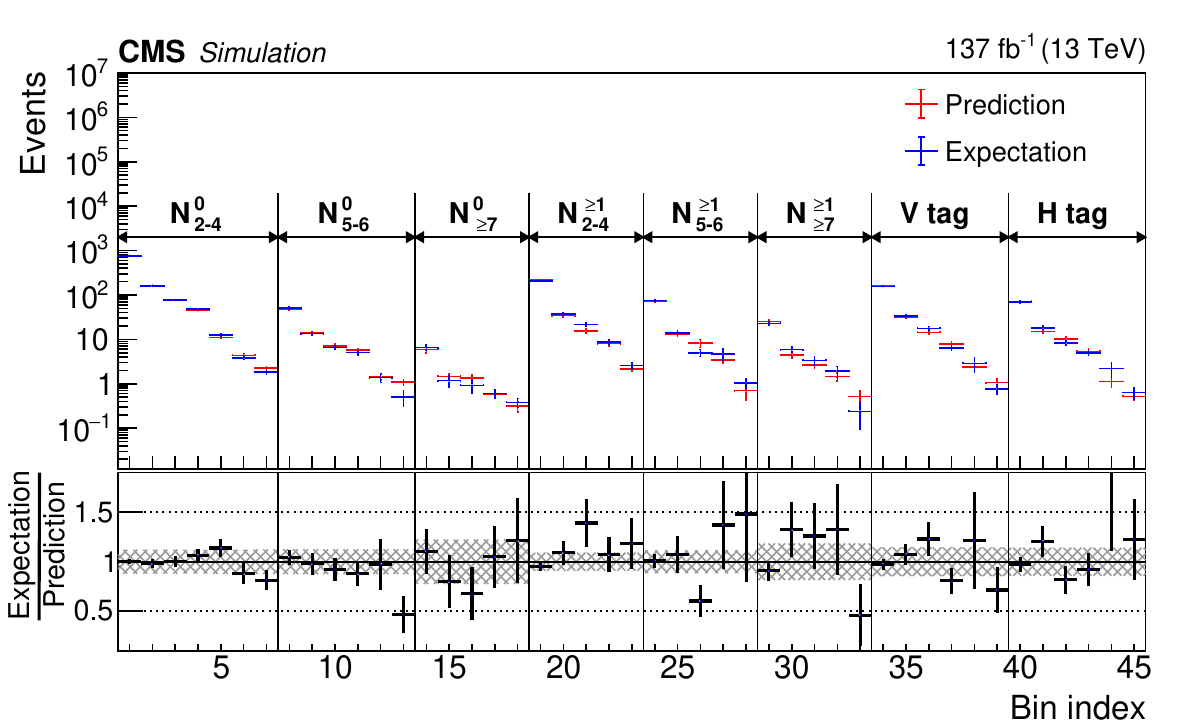}
  \caption{Left: the relative contributions of events with light lepton(s) or \tauh candidate(s) in the SRs and CRs (upper panel), and the corresponding transfer factors (TFs; see text), along with their statistical uncertainties (lower panel).
  Right: a comparison between the expected and predicted lost lepton event yields from simulated \wgammaj and \ttgammaj processes in each of the SR bins. The vertical error bars indicate the statistical uncertainty in the simulation and the hashed bands in the lower panel indicate the systematic uncertainties.}
  \label{fig:LLcontr}
\end{figure*}

An \Pe or a \PGm is considered ``lost'' when it fails reconstruction, identification, or isolation, or if it is outside the detector or kinematic acceptance, as described in Section~\ref{sec:det-reco}.
Events containing a \tauh candidate that is not rejected by the isolated track veto also contribute to the background, as the \tauh candidate is reconstructed as a jet.
We estimate both these contributions together as the lost-lepton background, given their similar origins.
The CRs used to estimate this background are collected by the same \ptmiss triggers as used for the SR events. There is one CR corresponding to each SR.
Except the \Pe, \PGm, and isolated track veto, the CRs are required to satisfy all criteria used to define the SRs.
Instead, we require the presence of exactly one electron or one muon, and no additional isolated tracks in the event.
It is important to note explicitly that these CR events are required to have a reconstructed photon in the final state.
We require $\mT(\pt^{\Pell}, \ptmiss) < 100\GeV$ to veto events potentially arising from new physics in similar single-lepton final states.
The relative composition of \wgammaj and \ttgammaj events in the single-\Pe and single-\PGm CRs, in the bins of \nj and \nb for SP SRs and \vandhtags for EW SRs, is shown in Fig.~\ref{fig:LLcontr} (left).
The SR events are shown as events containing a lost \Pe, a lost \PGm, or a \tauh candidate.
Since we do not use a dedicated \tauh veto, the last type of events are a large fraction of the lost-lepton background for this search.

The background estimation makes use of transfer factors (TFs) from the simulation. 
The TFs are defined as the ratio of the number of events in the SRs to the number of events in the respective CRs. 
The simulated event yields are corrected for known differences with data in lepton identification, \PQb tagging, and trigger efficiency.
The TFs are calculated in the bins of \nj and \nb for the SP SRs and in the bins of \vandhtag for the EW SRs; they are not binned in \ptmiss and include events with $\ptmiss>200\GeV$.
The number of events predicted in SR bin $i$, $\NdataSRlep(i)$, is obtained from the number of events in the corresponding CR, $\NdataCRlep(i)$, as
\begin{equation}
  \label{eq:LLTF}
  \NdataSRlep (i) = \text{TF}(\nj,\nb ~\text{or}~\vtag, \htag) \NdataCRlep(i).
\end{equation}
Here, the TFs are applied to the single \Pe and \PGm data CRs as per-event weights, depending on the event characteristics in terms of \nj and \nb for the SP SRs and \vorhtag for the EW SRs. 
As shown in the lower panel of Fig.~\ref{fig:LLcontr} (left), these TFs vary from 0.5 to 1.0, depending on the SR.

We validate the background estimation method and the applicability of these TFs in the simulation by using the 1\Pe\PGg and 1\PGm\PGg CRs to predict the events in the SRs using Eq.~\eqref{eq:LLTF}. 
Comparisons of the expected and predicted event yields in the SRs are presented in Fig.~\ref{fig:LLcontr} (right).
The discrepancies between the expected and predicted yields, especially in the high \ptmiss or high \nj bins, are covered by the statistical uncertainties.

The predicted number of lost-lepton background events can be affected by several sources of uncertainty that impact the TFs.
To account for potential mismodeling of collinear photon radiation from quarks and leptons, the cross sections of relevant simulated samples are varied by 20\%~\cite{CMS:2021klw} to account for differences in the modeling of photon radiation in different simulated samples and the effect on the TFs is found to be 4\%.
The uncertainty in the PDFs is evaluated by varying event weights from 100 PDF replicas~\cite{NNPDF:2014otw,Ball:2017nwa}, resulting in a 3\% effect.
To evaluate the renormalization ($\muR$) and factorization ($\muF$) scale uncertainties, different weights obtained by varying the scales independently by 0.5 and 2~\cite{Cacciari:2003fi,Catani:2003zt} are used. 
The effect of this uncertainty is found to be 2\% on the predicted number of lost-lepton background events. 
The uncertainties in the corrections applied to the simulation are propagated by varying their values and recalculating the predicted event yields in each SR.
The variation in the number of predicted event yields with respect to the central values is taken as the respective uncertainty for each correction.
The uncertainties in the lepton identification efficiency, \PQb tagging efficiency, and the JEC and JER lead to 0.6, 0.7, 6, and 6\% uncertainties in the lost-lepton background, respectively.
The limited size of the simulated samples results in a 2--10\% uncertainty in the predicted event yields. 
The estimated numbers of lost-lepton background events in each SR, along with uncertainties, are shown in Section~\ref{sec:results}.

\subsection{Misidentification of electrons as photons}
\label{subsec:efake}
If the track associated with an electron is not reconstructed or linked to it, the energy deposited by the electron in the ECAL could potentially be misidentified as a photon.
Such electrons typically arise from \wjets, \ttjets, and single top quark events in which a \PW boson decays to an electron and a neutrino.
To estimate the background from electrons misidentified as photons, CRs are defined to include events containing exactly one electron with $\pt > 100\GeV$ and zero photons satisfying the criteria described in Section~\ref{sec:det-reco}. 
Similar to the lost-lepton CRs, we require the \mT of the electron and \ptmiss to be $<100\GeV$.
Jets that have $\dR < 0.3$ with respect to the selected electron are not considered when computing \ST and \nj.

The misidentification rate, $f$, is defined as the ratio of the number of events with a misidentified electron to the number of single-electron events.
The rate $f$ is determined using a sample of simulated \wjets and \ttjets events containing $\PW\to\Pe\PGn$ decays. 
It depends on the kinematic properties of the electron and the presence of jets and other particles near the electron.
The activity around the electron is characterized using the charged multiplicity, denoted \Qmult and defined as the number of charged constituents of the closest jet to the electron or photon with $\dR(\text{jet}, \Pe~\text{or}~\PGg) < 0.3$.

A correction factor $\alpha$ is included to account for differences in the rate $f$ between data and simulation.
The factor $\alpha$ is obtained from data and simulated events containing an \EE pair with the ``tag and probe'' method~\cite{CMS:2010svw}.
The tag electron is required to have $\pt > 40\GeV$ and to match within $\dR<0.2$ with a generator-level electron arising from a \PZ boson in simulation or a trigger-level electron object in data.
The probe electron (photon) is selected with requirements similar to the electron in the 1\Pe CR (SR). 
The pair of the tag electron and the probe electron or photon is also required to satisfy $\dR(\rm{tag}, \rm{probe})>0.2$ and to have an invariant mass within 80--100\GeV.
In this case, the rate $f$ is defined as the ratio of the number of events with a photon as the probe to the number of events with an electron as the probe. 
The factor $\alpha$ is determined as the ratio of the rate $f$ measured in data and simulated events, evaluated separately for events with \nb = 0 and $\geq$1.
To account for different run conditions, the factor $\alpha$ is also measured separately for data recorded in 2016, 2017, and 2018.
The values of $\alpha$ vary from 1.9--2.4.

\begin{figure}[h!]
  \centering
  \includegraphics[width=0.68\linewidth]{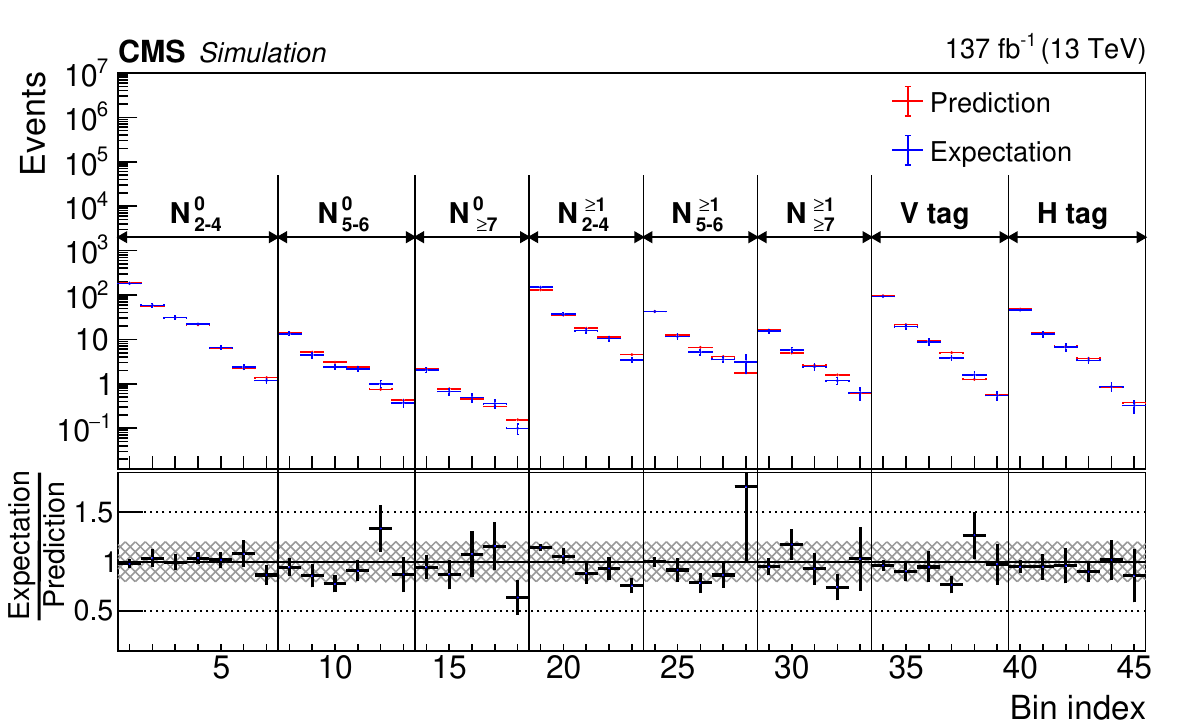}
  \caption{A comparison of the number of events with an electron misidentified as a photon in the SRs and the number estimated using the single-electron CRs with simulated samples. The ratio of the expected and predicted event yields in each SR is shown in the lower panel. The shaded region in the lower panel indicates the systematic uncertainties in the predicted number of background events.}
  \label{fig:FRclosure}
\end{figure}

The number of events with the electrons misidentified as photons in the SR bin $i$, $\NdataSRefake(i)$, is estimated from the number of events in the corresponding CR, $\NdataCRefake(i)$, as
\begin{equation}
  \label{eq:fake}
  \NdataSRefake(i) = f(\pt^{\Pe}, \Qmult) \alpha(\nb) \NdataCRefake(i).
\end{equation}
For a given CR, the rate $f$ is applied as a per-event weight according to the \pt of the electron and the value of \Qmult for the jet closest to the electron.
A comparison of the number of events expected in the SRs and predicted by the single-electron CRs, with both the SRs and CRs taken from simulated \wjets and \ttjets events, is shown in Fig.~\ref{fig:FRclosure}.

The uncertainties in the \PQb tagging efficiencies, JECs, and JER smearing are propagated to the simulated events used to determine the rate $f$, and their effects on the final predicted event yields are found to be ${<}1$, 3, and 4\%, respectively.
The overall systematic uncertainty in $\alpha$ from topological differences in \zee compared to \ttbar and \wjets is estimated to be 20\%,
based on comparing simulated $f$ values in $\Pep\Pem$ events for the tag-and-probe and single-electron or single-photon selections.
This component dominates the total systematic uncertainty in the predicted number of events with electrons misidentified as photons and is taken to be correlated across all SR bins.
The uncertainties arising from the limited number of CR and SR events in the simulated samples are up to 20\%.
The statistical uncertainty in the single-electron CR in data contributes up to a 20\% uncertainty in the predicted number of background events in the SRs.
The background prediction in the SRs, along with the uncertainties, is shown in Section~\ref{sec:results}.

\subsection{\texorpdfstring{\znunugammaj}{Z(nu nu)photon+jets} background}
\label{subsec:zinv}

The presence of energetic neutrinos in \znunugammaj events manifests as large \ptmiss and results in significant background for searches in final states requiring zero leptons, particularly in low-\nj and high-\ptmiss SRs.
We use \znunugammaj simulated events to estimate the predicted event yields in the SP SRs, which are defined by \nj, \nb, and \ptmiss. 
These event yields are adjusted by the ratio of the number of \zeegammaj and \zmumugammaj events, collectively called \zllgamma events, measured in data and simulation to account for any potential mismodeling of \zgammaj production. 
The leptonic final states of \PZ boson decays have limited numbers of events because of their small branching fractions and hence are not directly used to estimate the \znunugammaj background. 

\begin{figure}[h!]
  \centering
  \includegraphics[width=0.49\linewidth]{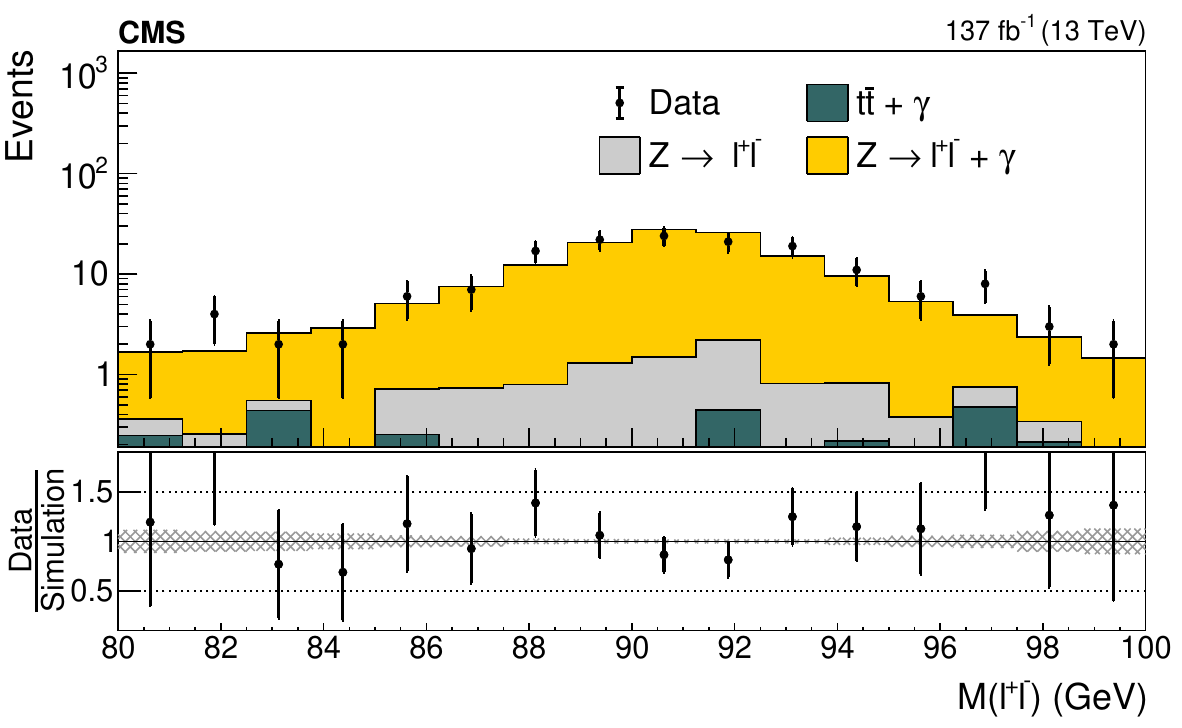}
  \includegraphics[width=0.49\linewidth]{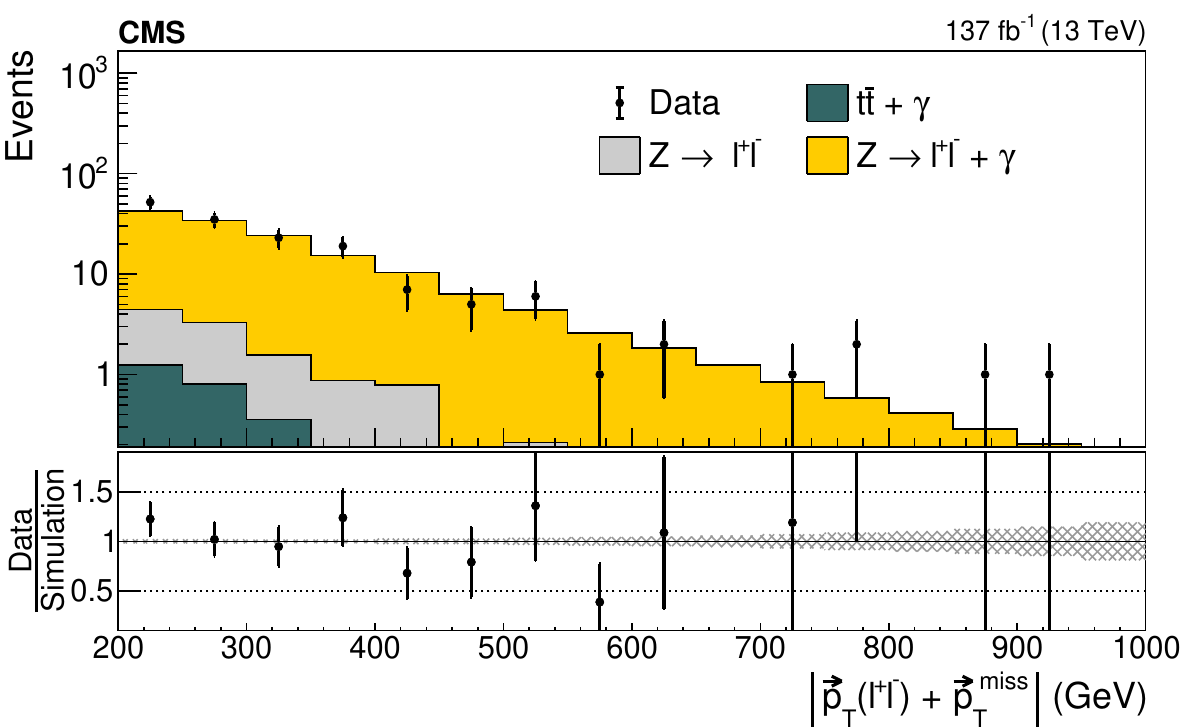}
  \caption{The distributions of the dilepton invariant mass (left) and the magnitude of the dilepton \ptvec plus the \ptvecmiss (right) for \llg events in data and simulation. The error bars represent the statistical uncertainty in the data events. In the lower panel, the shaded region shows the statistical uncertainty in the simulation.}
  \label{fig:ZinMassMET}
\end{figure}

To select \zllgamma events, we require a pair of light leptons with the same flavor, opposite charge, and invariant mass in the range of 80--100\GeV, depicted in Fig.~\ref{fig:ZinMassMET} (left).
The \ptmiss in these events is required to be less than 200\GeV. However, to mimic the kinematic properties of the \znunu process in the SRs, the dilepton system should be treated as invisible.
Therefore, the magnitude of the sum of the dilepton \ptvec and the \ptvecmiss, shown in Fig.~\ref{fig:ZinMassMET} (right), must be greater than 300\GeV, following the baseline selection.
We also require $\dR(\Pell,\PGg) > 0.2$ to ensure that the photon is not radiated from one of the leptons.
The numbers of events obtained in data and simulation using these criteria are denoted \NdataCRzll and \NmcCRzll, respectively. Here, the subscript $\Pell$ refers to electrons and muons. 

The number of events in the SRs is obtained using the expression:
\begin{equation}
\NdataSRznunu(i) = \left(\frac{\NdataCRzll \betazll}{\NmcCRzll}\right)_{j} \NmcSRznunu(i),
\end{equation}
where $\NdataSRznunu(i)$ and $\NmcSRznunu(i)$ are the numbers of events predicted in SR $i$ in data and simulation, respectively, and
$\betazll = 1 - N_{\ttgamma}/N_{\llg}$ is a correction factor to account for the contribution of \ttgammaj processes to the $\llg$ events in data. The index $j$ = 0 or 1 corresponds to \nb = 0 or $\geq$1.
The contribution of \ttgamma is estimated from simulation and it is found to be statistically compatible with the number of opposite-sign, different-flavor ($\Pe^{\pm}\mu^{\mp}$) events from data.
As the contamination of \ttgamma events is expected to be higher in the SRs with \nb $\geq$1, the factor \betazll is derived separately for the regions with and without a \PQb-tagged jet. 

The corrections to \znunugammaj events from $\llg$ events are $1.07\pm0.09$ and $1.01\pm0.28$ for \nb = 0 and $\geq$1, respectively.
These uncertainties are propagated to the predicted number of events in the SRs along with the statistical uncertainty in the \znunugammaj simulated samples, which ranges from 2--70\%.
In both the SP and EW SRs, the statistical uncertainty is larger in the high-\ptmiss bins.
To account for any mismodeling of photon \pt in the simulation, we apply an additional systematic uncertainty, which ranges 18--40\%, depending on the \ptmiss bin. This is assessed based on the results presented in Ref.~\cite{Denner:2015fca}.
The number of \znunugammaj events for each SR, along with the uncertainty in the prediction, is presented in Section~\ref{sec:results}.

\subsection{\texorpdfstring{\gjets}{photon+jets} and QCD multijet background}
\label{subsec:multijet}

QCD multijet and \gjets events with a well-reconstructed photon can contribute to the SRs if they also contain large \ptmiss resulting from significant mismeasurement of one or more jets or from the presence of neutrinos from the semileptonic decays of heavy-flavor hadrons.
We use an ``ABCD'' method to estimate this background, where the regions A, B, and C are data CRs designed to be nonoverlapping with the SRs and the region D corresponds to the SRs. 
To define these CRs, we use events with $200 < \ptmiss < 300 \GeV$ (low-\ptmiss) or $\dphifull < 0.3$, indicating at least one of the two leading jets is aligned with the direction of \ptmiss (low-\dphi).
These events are required to satisfy all other baseline selection criteria.
From the observed event yields in every CR, we subtract the contributions from the lost-lepton, misidentified electron, and \zgammaj backgrounds, which are obtained using the same methods as used for the SRs.

The CRs A and C consist of events with low-\ptmiss, with the former being low-\dphi and the latter being high-\dphi.
These events are divided into the same exclusive regions based on \nj, \nb, \vandhtag as used to define the SP and EW SRs.
In each of these regions, a ratio \Rlowmet is defined as the number of events with high-\dphi (CR C) to that with low-\dphi (CR A).
The CR B consists of events with high-\ptmiss and low-\dphi. These events are divided into the same regions as the SRs, including the \ptmiss binning.
The predicted number of events in SR $i$ is then:
\begin{equation}
  \label{eq:qcd}
  N^{\rm{data}}_{\mathrm{multijet}+\gamma} (i) = \Rlowmet(r) \kappa (r) N^{\rm{data}}_B (i),
\end{equation}
where the index $i$ refers to the SR and the index $r$ refers to the bins defined based on \nj, \nb, \vtag, and \htag as used for the SRs.
The factor $\kappa (r)$ is used as a correction determined from the simulation to account for any differences in the ratios of the low-\ptmiss regions A and C compared to the high-\ptmiss regions B and D.
It is calculated in the same bins as used for \Rlowmet as
\begin{equation}
  \label{eq:kappa}
  \kappa(r) = \frac{\Rhighmet^{\text{MC}}(r)}{\Rlowmet^{\text{MC}}(r)}.
\end{equation}

This method is validated in the simulation. The CRs and SRs are defined using the simulated \gjets and QCD multijet events. The same event sample is also used to measure \Rlowmet. The numbers of events predicted by the method in various SRs are found to be consistent with the expected values, within the statistical precision.
The test of the method in simulation is nontrivial because the parametrization of $\kappa$ is based on \nj, \nb, \vandhtag bins, whereas the predictions are done for each of the SRs, which also include \ptmiss binning.

\begin{figure}[h!]
  \centering
  \includegraphics[width=0.68\linewidth]{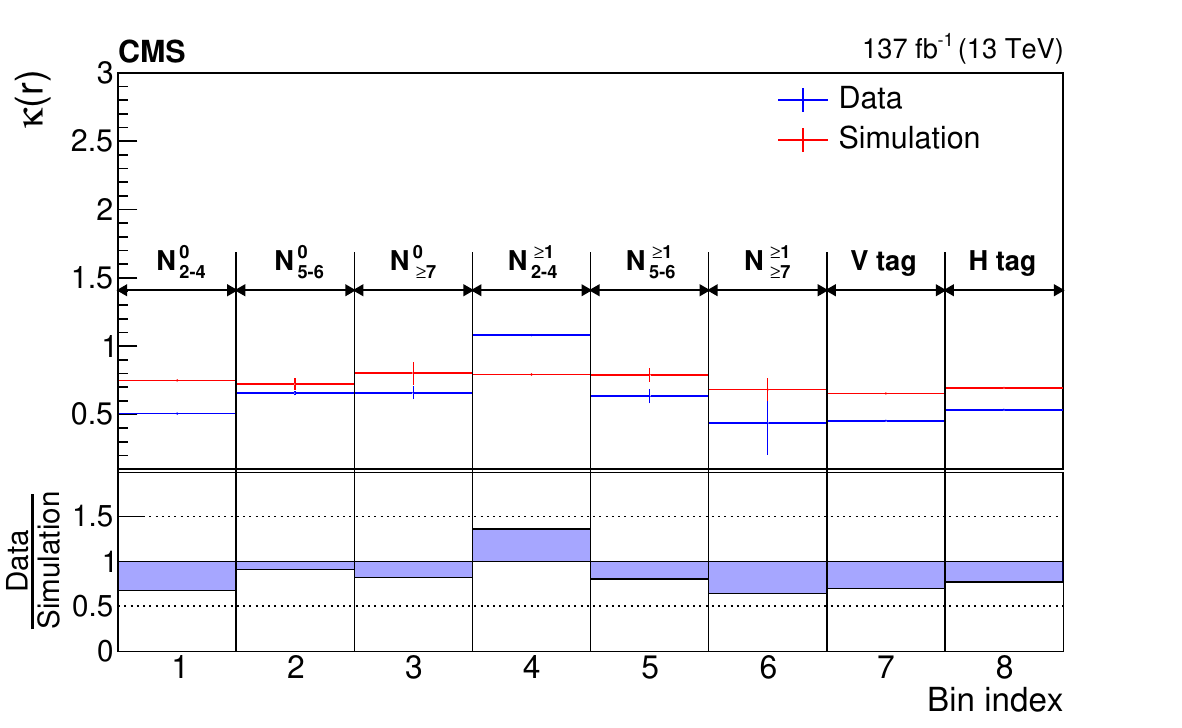}
  \caption{A comparison between $\kappa$ estimated from simulation and from data in the zero-photon CR. The values are given for each \nj, \nb, \vtag, and \htag bin, represented as $r$. The blue bands in the lower panel represent the relative systematic uncertainty in $\kappa$.}
  \label{fig:kappa_data_mc}
\end{figure}

Since $\kappa$ is an important ingredient for this background estimation and it is obtained from the simulation, we validate these factors in data and simulation samples using events that do not contain a photon, referred to as zero-photon events. 
The zero-photon region is dominated by multijet events.
The contributions from the other SM background processes are small, so they are estimated from the simulation and subtracted from the zero-photon event yields.
The values of $\kappa$ derived from data and simulation differ by 10--36\%, as shown in Fig.~\ref{fig:kappa_data_mc}. 
The observed differences are treated as systematic uncertainties in the predicted number of events.
There is one $\kappa$ value for each set of \nj, \nb, \vtag, and \htag requirements,
so the uncertainty in $\kappa$ is correlated among all the corresponding SRs.
The statistical uncertainties in $\kappa$ and the number of events in low-\ptmiss regions vary in the range 10--50\%. 
Section~\ref{sec:results} summarizes the predicted number of \gjets and QCD multijet events in the SRs.

\section{Systematic uncertainties}
\label{sec:syst}
The sources of systematic uncertainty and their effects on the predicted numbers of events for the lost-lepton, electron misidentified as photon, \znunuj, and \gjets backgrounds have been discussed in Section~\ref{sec:smbkg}.
The following sources of systematic uncertainty are considered for the simulated signal event yields. 
The uncertainties in the total integrated luminosity~\cite{CMS-LUM-17-003,CMS-PAS-LUM-17-004,CMS-PAS-LUM-18-002} affect the signal yield in each SR. This effect is taken to be 1.6\% for all the SRs and all signal models.
The uncertainties in the JEC and JER measurements from data are propagated to individual jets, and their effect on event yields is taken as an uncertainty in the respective SR bins. 
The effect of variations in the pileup reweighting is estimated in a similar way. 
The uncertainties related to pileup, JECs, and JER contribute approximately 2\% each.
The uncertainty in the trigger efficiency for simulated signal events varies from 3--10\%, with larger values for SRs with lower \ptmiss.
The uncertainties related to the isolated track veto and jet ID modeling in the fast simulation are 2 and 1\%, respectively.
The \PQb tagging efficiency and light flavor quark mistagging rates in the fast simulation signal samples are corrected with factors derived from the \GEANTfour-based simulation and data. 
The uncertainties in these corrections are propagated to the final signal yield, and their effect in the SP SRs is up to 10\%.
The statistical uncertainty from the limited number of simulated events in the signal samples ranges 0.7--38\% when considering all bins;
when considering only the most sensitive bins, this uncertainty decreases to 0.1--7\%.

The PDF uncertainty, and the $\muR$ and $\muF$ scale uncertainties that affect the total production cross section are treated as theoretical uncertainties. The effects of these on the signal yield are 2 and 5\%, respectively.
The modeling of initial-state radiation (ISR) in the signal simulation is corrected by applying data-to-simulation correction factors~\cite{CMS:2019zmd}.
The corrections depend on the number of ISR jets and the \pt of the chargino-neutralino system for the strong production and the electroweak signal models, respectively.
These uncertainties have magnitudes of 25\% for strong production and 10\% for electroweak signal models.
Table~\ref{tab:syst} summarizes the sources of systematic uncertainty and their effects on the predicted backgrounds and signal yields.

\begin{table}[h!]
  \centering
  \topcaption{The systematic uncertainties in the predicted background and signal event yields (in \%). A dash (\NA) indicates that the source of uncertainty is not applicable or negligible.}
  \cmsTable
  {
  \label{tab:syst}
  \begin{tabular}{lccccc}
    Source		& Lost lepton & Misidentified \Pe & \zgamma &	Multijet$+\PGg$ &  Signal \\ \hline
    Integrated luminosity	        & \NA	      & \NA		  & \NA     &   \NA		  &  1.6 \\
    Limited number of CR events & 3--100      & 5--20	  & 8--28   &	2--100		  &  \NA \\
    Limited number of simulated events & 2--10	      & 2--20	  & 2--70   &	10--50 &  0.7--38 \\
    \PQb tagging	& 0--1	      & 0--1		  & \NA	    &	\NA		  &  0--10 \\
    PDF			&	3     & \NA		  & \NA	    &	\NA		  &  1--2 \\
    $\muR$ and $\muF$ scales & 2     & \NA		  & \NA	    &	\NA		  &  0.3--5 \\
    JEC			&   0--6	      & 0--3		  & \NA	    &	\NA		  &  1--2 \\
    JER			&   0--6	      & 0--4		  & \NA	    &	\NA		  &  1--2 \\
    Pileup		&   \NA       & \NA	  & \NA	    &	\NA		  &  0.1-0.3 \\
    Trigger efficiency &   \NA       & \NA	  & \NA	    &	\NA		  &  3--10 \\
    Collinear $\gamma$  &   4	      & \NA		  & \NA	    &	\NA		  &  \NA	\\	
    $\alpha$ 		&   \NA	      & 20		  & \NA	    &	\NA		  &  \NA	\\
    Modeling of $\gamma$ \pt & \NA       & \NA		  & 18--40  &	\NA		  &  \NA	\\
    $\kappa$ modeling	&   \NA	      & \NA		  & \NA	    &	10--36		  &  \NA	\\
    Stat. unc. in low-\ptmiss A, C regions	&   \NA	      & \NA		  & \NA	    &	10--50	          &  \NA	\\
    Isolated track veto & \NA    & \NA		  & \NA	    &	\NA		  &  2  \\
    Jet ID    & \NA    & \NA		  & \NA	    &	\NA		  &  1 \\
\end{tabular}
}
\end{table}

\section{Results and interpretation}
\label{sec:results}

\begin{figure*}[h!]
\centering
\includegraphics[width=0.98\linewidth]{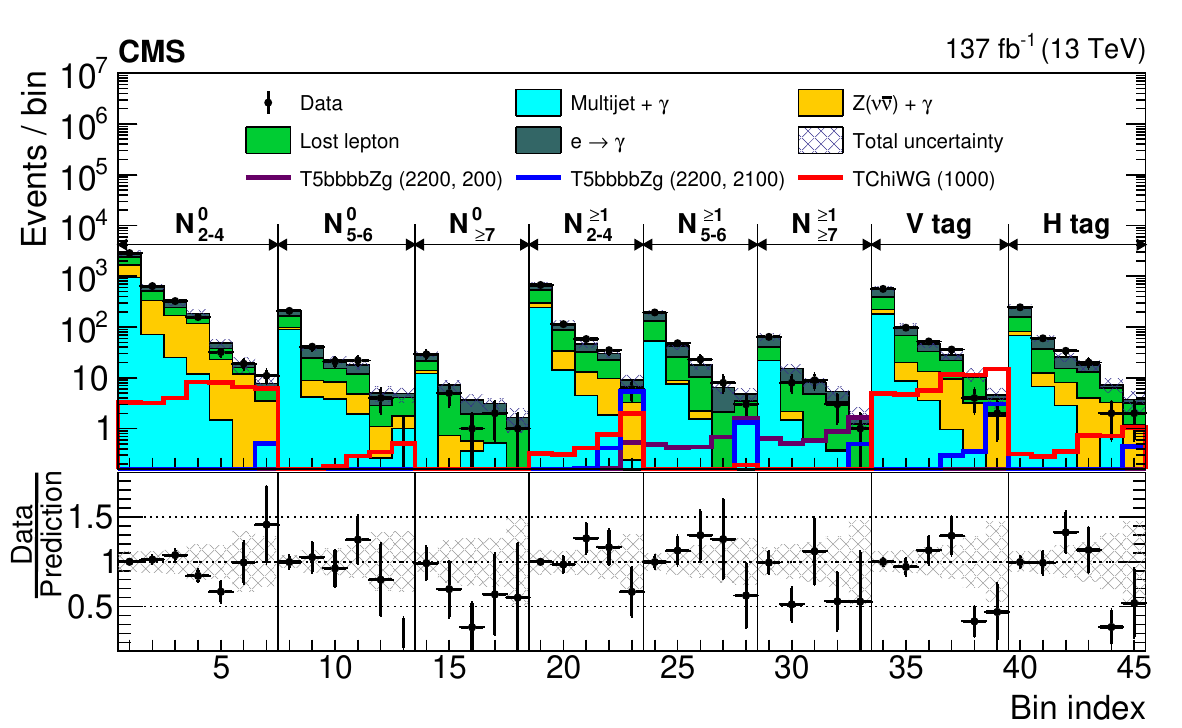}
\caption{The numbers of predicted background events and observed events in the SRs and low-\ptmiss CRs. The lost-lepton, electron misidentified as photon, \zgammaj, and \gjets and QCD multijet backgrounds are stacked histograms. The observed numbers of events in data are presented as black points. For illustration, the expected event yields are presented for the signal model T5bbbbZG, for small (blue) and large (purple) differences in the masses of the \PSg and NLSP. Also shown is the expected distribution of events for the signal model TChiWG (red). The numerical values in parentheses in the legend entries for the signal models indicate the \PSg and NLSP mass values in {\GeVns} for strong production and the NLSP mass value for electroweak production. The lower panel shows the ratio of the observed number of data events and the predicted backgrounds. The error bars represent the statistical uncertainty in the data events, and the shaded band represents the statistical and systematic uncertainties in the predicted background. The \ptmiss bin 200--300\GeV is used for the estimation of the $\gjets$ and QCD multijet background.}
\label{fig:TotPredBkg}
\end{figure*}

We perform a simultaneous maximum likelihood fit to the number of events in the low- and high-\dphi regions to predict the number of SM background events in the SRs.
The profile likelihood ratio \qmu is used as the test statistic to compute limits
in the modified frequentist \CLs approach~\cite{Junk:1999kv,Read:2002hq}, employing the asymptotic approximation~\cite{Cowan:2010js}.
The statistic is defined as $\qmu =  - 2 \ln(\Lmu/\Lmax)$, where \muSig is the SUSY signal strength,
\Lmax is the maximum likelihood from varying all parameters including \muSig, and \Lmu is the maximum likelihood for a fixed \muSig.
The observed numbers of events in various CRs are modeled using gamma distributions, which correctly represent the statistical uncertainties.
The predicted yield of signal events in each CR is found to be negligible.
The other systematic uncertainties listed in Table~\ref{tab:syst} are modeled as log-normal constraints in the likelihood.
The results obtained from the CR-only fit under the background-only hypothesis are shown in Fig.~\ref{fig:TotPredBkg}.
The numerical values, including the uncertainties in each background prediction, are given in Appendix~\ref{app:table}.
In most of the SRs, the observed event yields are consistent with the predictions, indicating no significant presence of signal events.
The maximum deviation observed is about 2 standard deviations below the prediction, in bin 13 ($5 \leq \nj \leq 6$, $\nb = 0$, $\ptmiss \geq 750\GeV$), bin 16 ($\nj \geq 7$, $\nb = 0$, $370<\ptmiss<450\GeV$), and bin 44 (\htag, $600<\ptmiss<750\GeV$).

\begin{figure*}[htb!]
\centering
\includegraphics[width=0.49\linewidth]{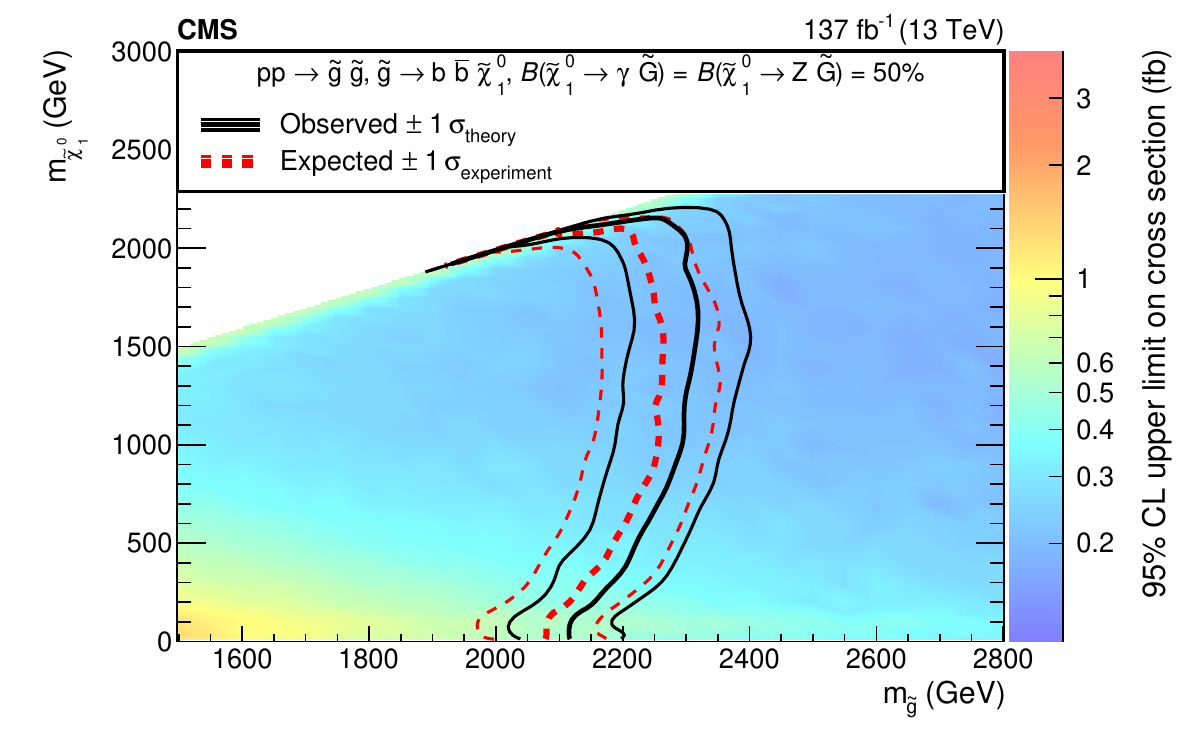}
\includegraphics[width=0.49\linewidth]{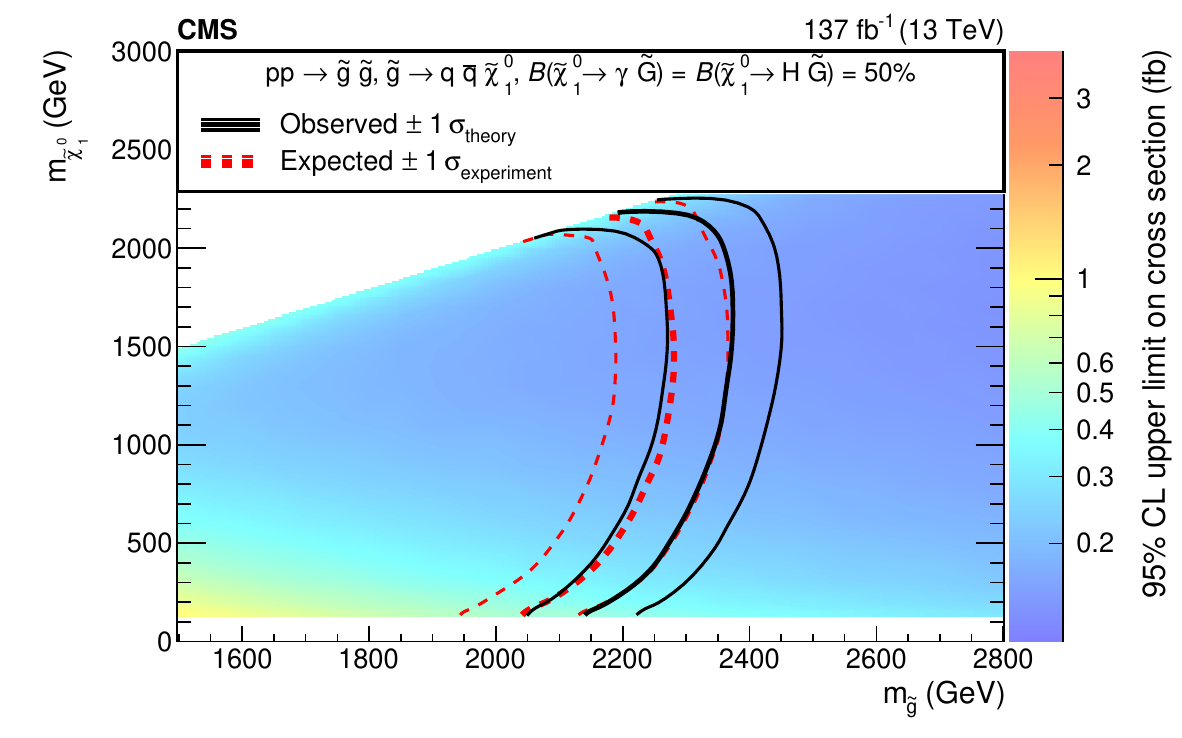}\\
\includegraphics[width=0.49\linewidth]{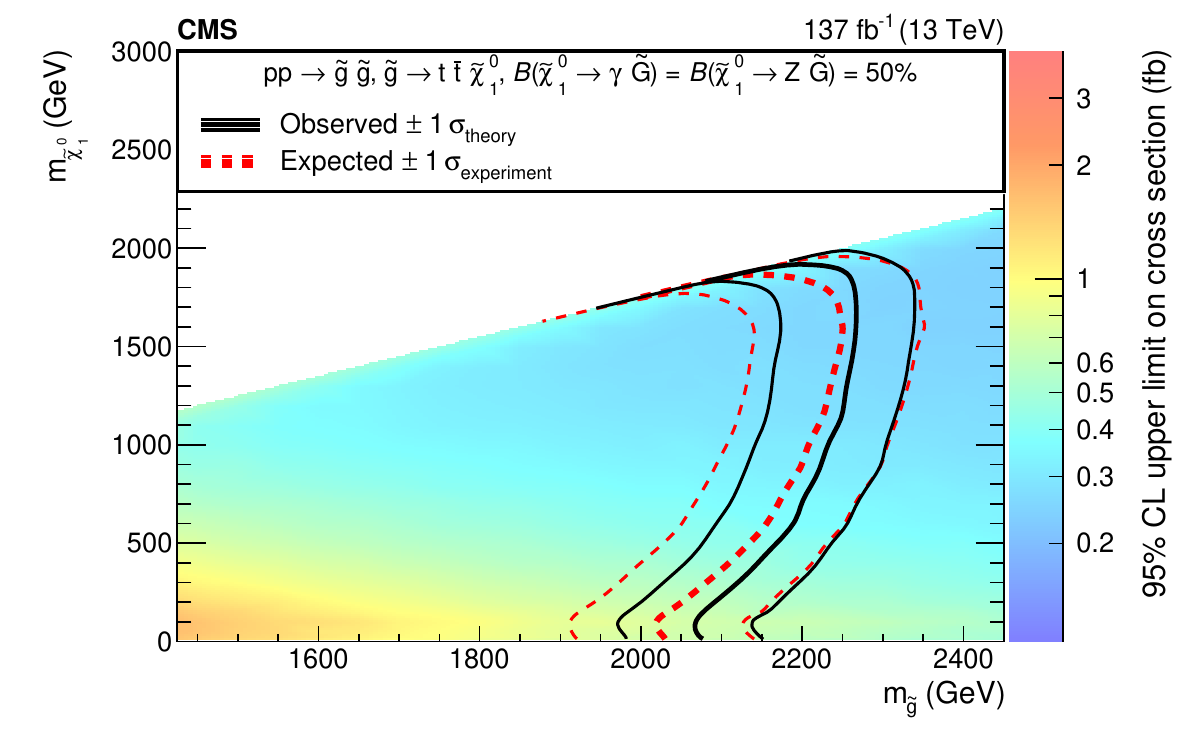}
\includegraphics[width=0.49\linewidth]{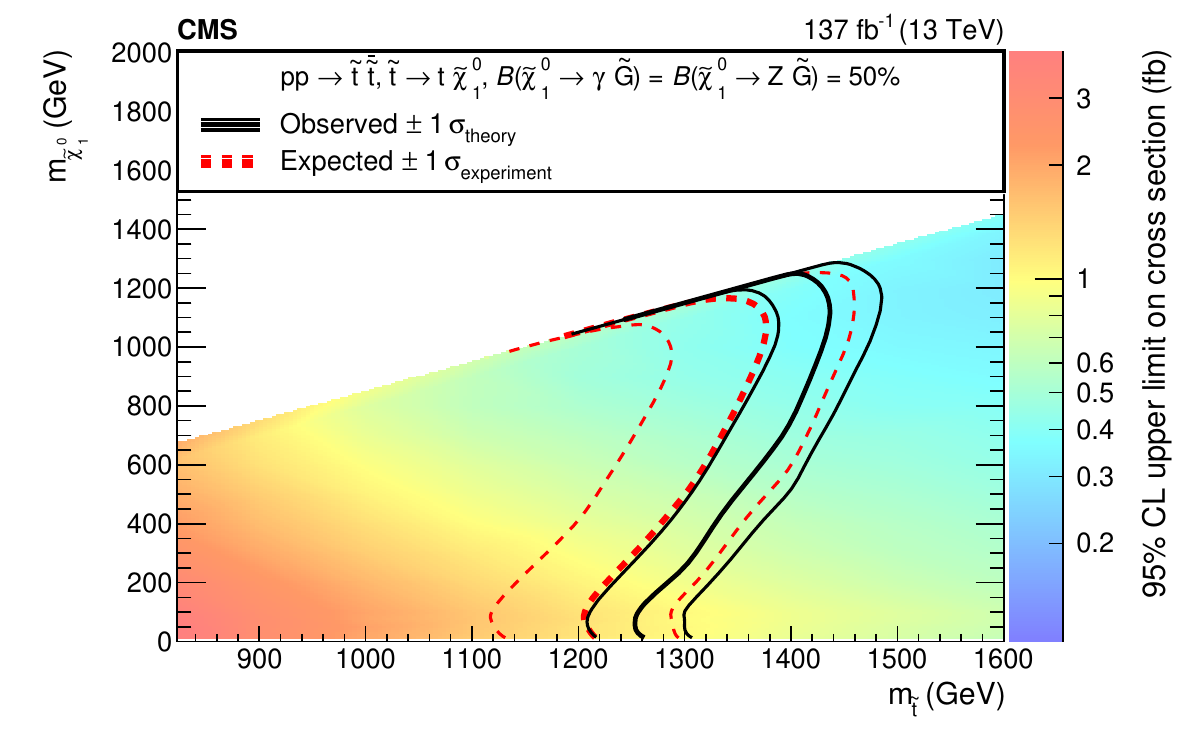}
\caption{The 95\% \CL upper limits on the production cross sections for \PSg pairs, with \GluinoToBBChiNOne followed by \ChiNOneToZsGra or \ChiNOneTogsGra (upper left, T5bbbbZG model), \GluinoToQQChiNOne followed by \ChiNOneToHsGra or \ChiNOneTogsGra (upper right, T5qqqqHG model), \GluinoToTTChiNOne followed by \ChiNOneToZsGra or \ChiNOneTogsGra (lower left, T5ttttZG model), or top squark pairs assuming the top squark decays to a top quark and a \ChiNOne followed by \ChiNOneToZsGra or \ChiNOneTogsGra (lower right, T6ttZG model). The thick black curve represents the observed exclusion contour and the thin black curves show the effect of varying the signal cross section within the theoretical uncertainties by $\pm 1\sigma_{\text{theory}}$. The thick red curve indicates the expected exclusion contour and the thin red curves show the variations from $\pm 1\sigma_{\text{experiment}}$ uncertainties.}
\label{fig:Exlstrong}
\end{figure*}

\begin{figure*}[htb!]
\centering
\includegraphics[width=0.49\linewidth]{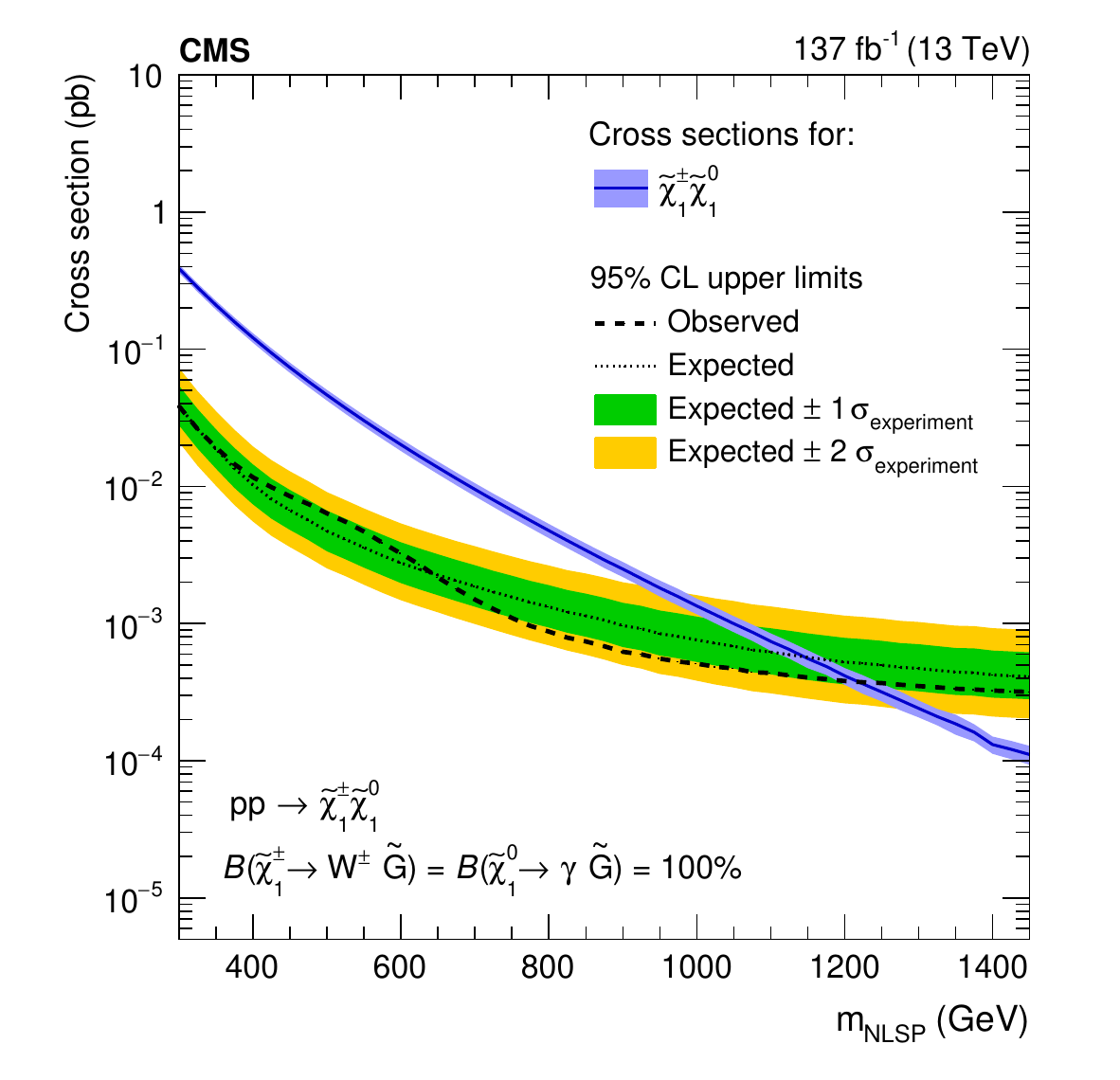}\\
\includegraphics[width=0.49\linewidth]{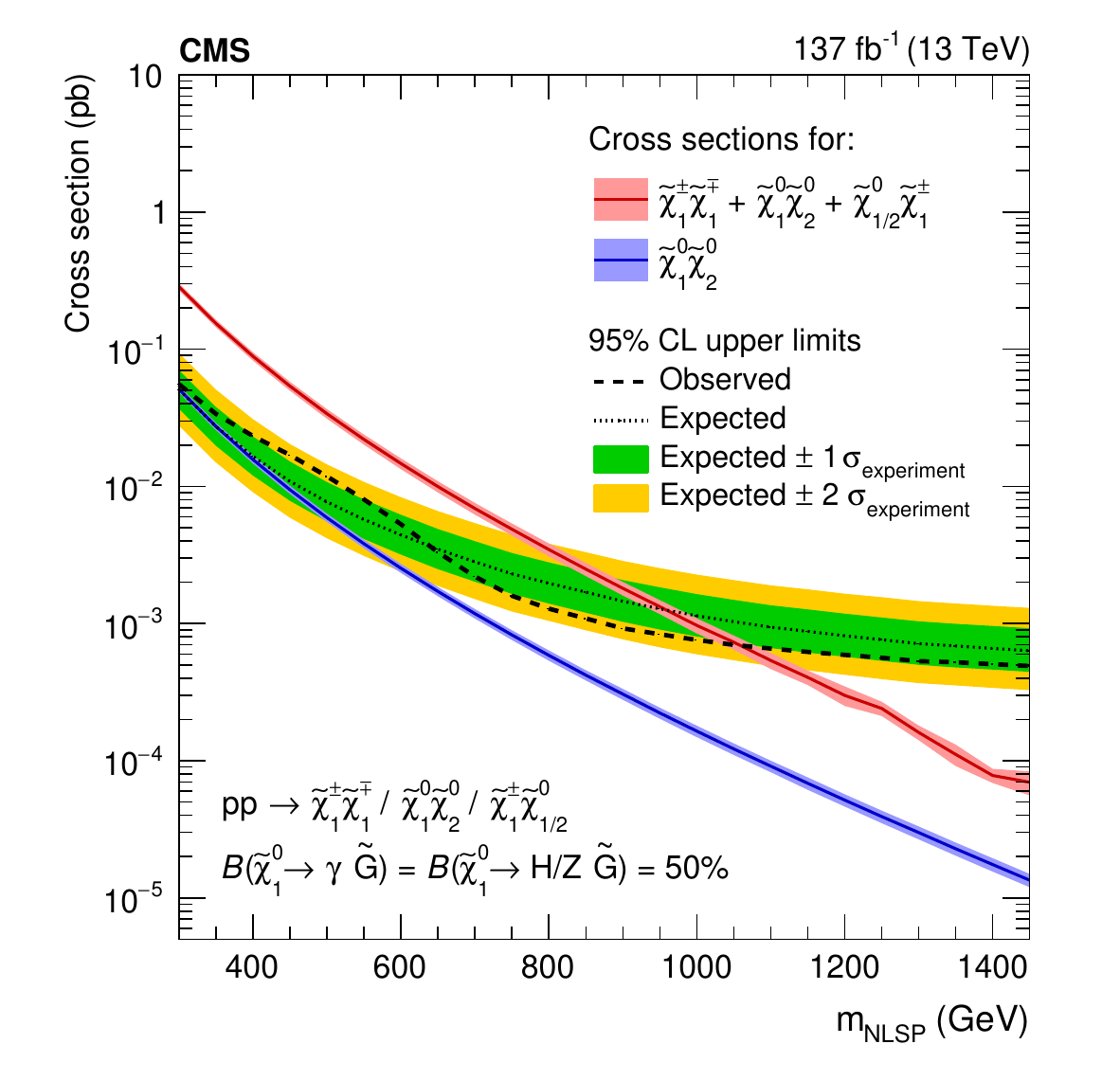}
\includegraphics[width=0.49\linewidth]{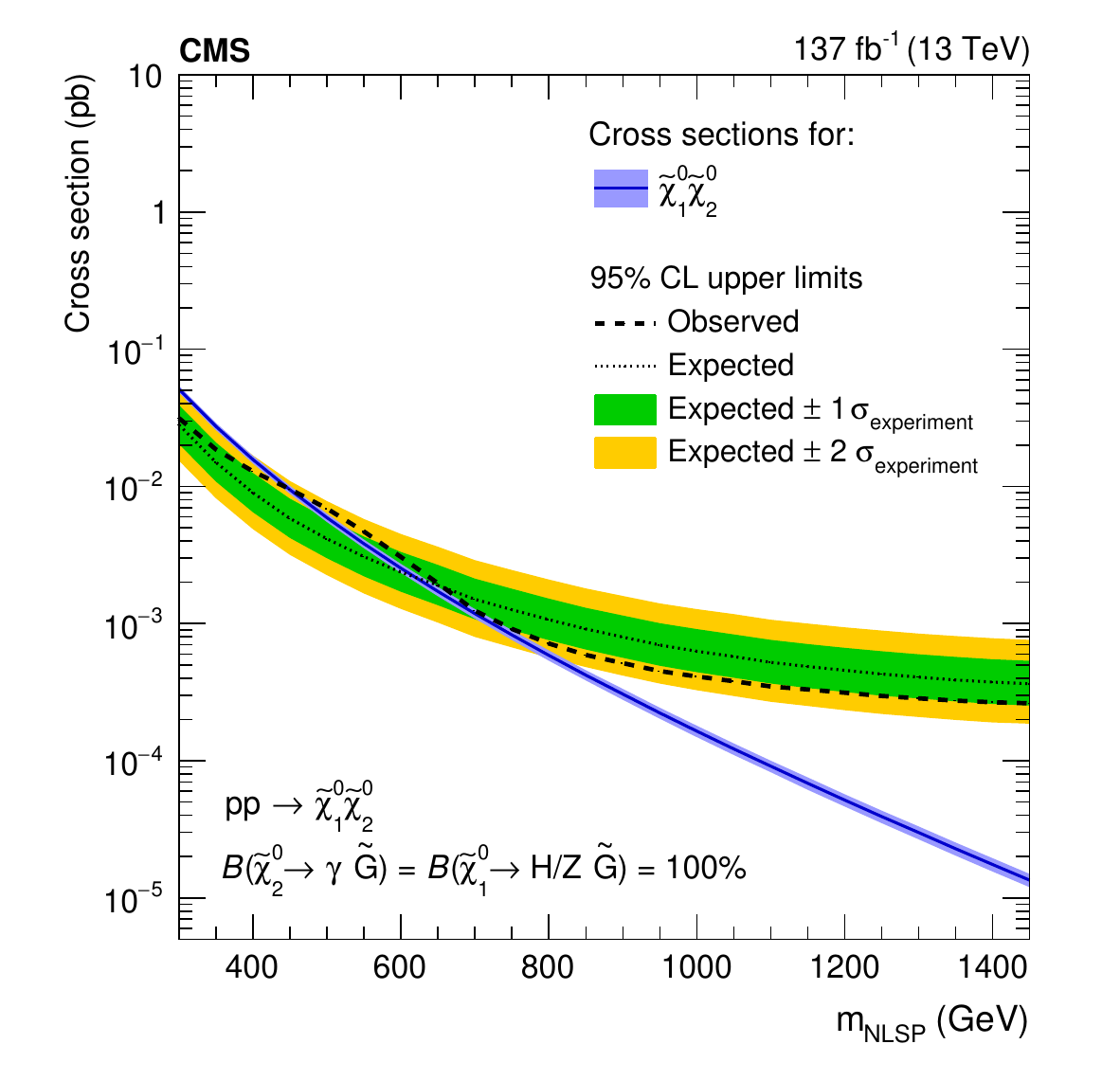}
\caption{The expected and observed limits on the electroweakino mass in the TChiWG (upper), TChiNG (lower left), and TChiNGnn (lower right) models at 95\% \CL. For the TChiNG model (lower left), scenarios with degenerate charginos and neutralinos leading to the combined process $\ChiCOneChiCOneMP+\ChiNOneChiNTwo+(\ChiNOne/\ChiNTwo)\ChiCOne$ (red) or the single process \ChiNOneChiNTwo (blue) are considered.}
\label{fig:ExlEWK}
\end{figure*}

The measured backgrounds along with their uncertainties and the observed number of events in the SRs are used to determine 95\% confidence level (\CL) upper limits on the production cross sections of various SUSY models, discussed in Section~\ref{sec:introduction}, using a maximum likelihood fit.
We compare these upper limits with theoretical production cross sections, and determine lower limits on masses of the SUSY particles in specific models and final states, which are excluded by this search. 
The exclusion limits in terms of masses of particles involved in a given model are shown in Fig.~\ref{fig:Exlstrong} for gluino and squark pair production scenarios.
Figure~\ref{fig:ExlEWK} presents the same for the production of electroweakino pairs.

For the gluino production model with a decay to \bbbar and NLSP (T5bbbbZG), the observed (expected) gluino mass exclusion is up to 2.32 (2.27)\TeV for small NLSP masses. In the T5qqqqHG and the T5ttttZG models, the observed (expected) upper limits on the gluino masses extend to 2.35 (2.30) and 2.26 (2.25)\TeV, respectively. 
The mass limits degrade for very high and very low NLSP masses. 
When the NLSP masses are large, the \ptmiss is large but the events contain lower hadronic activity (\nj). 
For the low NLSP masses, hadronic activity is high, but \ptmiss is low.
In these scenarios, either the signal acceptance is low or the signals populate SRs with larger backgrounds, resulting in a decrease in the sensitivity of the analysis. 
These features are illustrated in the open histograms shown in Fig.~\ref{fig:TotPredBkg}.
For the T5qqqqHG scenario, the observed limits are stronger than the expected ones because of the small deficit in the observed event yields in the high-\nj and $\nb=0$ regions and in the high-\ptmiss \htag regions.
In the strong production models that involve off-shell \PZ bosons and very low NLSP masses, the limits are stronger than those from on-shell \PZ bosons, because the former imparts larger \pt to the gravitinos, leading to larger \ptmiss.
In the T5qqqqHG model, off-shell Higgs boson decays are not considered.
For the top squark pair production model (T6ttZg), where the top squark decays into a top quark and NLSP, the expected mass limit is 1.38\TeV and the observed mass limit is 1.43\TeV.
There is an approximately 0.7 standard deviation difference between the expected and observed limits, which comes from signal regions with high \ptmiss, high \nj, and $\nb\geq1$.

In electroweak production models, for sufficiently large electroweakino masses, the signal events mostly populate large \ptmiss and \vorhtag signal regions.
In the TChiWG scenario, we observe (expect) the exclusion of \ChiCOneChiNOne masses up to 1.23 (1.17)\TeV, assuming wino-like \ChiCOneChiNOne production.
In the TChiNG model, the \ChiCOne, \ChiNOne, and \ChiNTwo are nearly mass degenerate, and we use higgsino-like cross sections to interpret the results. 
Electroweakino masses below 1.05 (0.95)\TeV are observed (expected) to be excluded, assuming the \ChiCOne and \ChiNTwo decays give rise to \ChiNOne and soft particles.
The expected upper limit on the mass is about 60--80\GeV smaller than the observed upper limit because of a deficit in the observed event yields in the highest \ptmiss bins of the \vandhtag SRs.
For the lower electroweakino masses, the signal populates moderate \ptmiss bins, which have observed event yields higher than the predictions by about one standard deviation. This leads to higher than predicted observed upper limits on the production cross section.
Alternatively, if only the \ChiNOneChiNTwo process occurs in this model, there is no exclusion in the range of NLSP masses considered in this search, based on the theoretically predicted cross section.
In the TChiNGnn model, with only the \ChiNOneChiNTwo process, the observed (expected) NLSP mass limit is 0.50 (0.65)\TeV.
The observed limit is weaker than expected because of upward fluctuations in several low-\nj $\nb\geq1$, \vtag, and \htag SR bins with intermediate \ptmiss values.

\section{Summary} \label{sec:summary}
A search for supersymmetry (SUSY) is presented using events with final states containing at least
one photon, large missing transverse momentum, and jets that may or may not arise from \cPqb quarks.
These signatures are motivated by models with gauge-mediated SUSY breaking (GMSB), in which the lightest SUSY particle (LSP) is a gravitino (\sGra)
and the next-to-LSP (NLSP) is a chargino (\ChiCOne) or neutralino (\ChiNOne), collectively called electroweakinos.
Several simplified models of strong production of pairs of gluinos (\PSg) and top squarks (\PSQt) are considered,
with the gluino decaying to a pair of quarks along with an NLSP or the top squark decaying to a top quark and an NLSP;
the NLSP then decays to a neutral gauge boson (photon, \PZ boson, or Higgs boson) and an LSP.
Models of pair production of electroweakinos are also considered, with the neutralinos decaying as described above,
and the charginos decaying to a \PW boson and an LSP.

Compared to previous searches, this search achieves increased sensitivity to scenarios with small mass differences between the gluino and the NLSP
with dedicated search regions based on identifying boosted massive bosons.
In addition, the search strategy is expanded to provide sensitivity to the production of electroweakino pairs.
The observations are consistent with the standard model expectations and 95\% confidence level upper limits
are set on the production cross sections of SUSY particles. In the
GMSB simplified models, the lower gluino mass limit reaches up to 2.35\TeV for models with $\GluinoToQQChiNOne$ followed by \ChiNOneToHsGra or \gsGra with equal probability, and
the top squark mass limit reaches up to 1.43\TeV for models with $\PSQt \to \cPqt \ChiNOne$ followed by \ChiNOneToZsGra or \gsGra with equal probability.
These results extend the previous mass limits~\cite{Sirunyan:2019vin} on gluinos and top squarks by 150--200\GeV.
For electroweakino pair production,
chargino and neutralino masses up to 1.23\TeV are excluded, assuming wino-like electroweakinos with decays \ChiCOneToWsGra and \ChiNOneTogsGra.
The higgsino-like electroweakino mass limits reach up to 1.05\TeV for models with \ChiNOneTogsGra, \ZsGra, or \HsGra with 50, 25, and 25\% branching fractions, respectively.
These are the best mass limits to date on electroweakino production with photons in the final state.

\begin{acknowledgments}
We congratulate our colleagues in the CERN accelerator departments for the excellent performance of the LHC and thank the technical and administrative staffs at CERN and at other CMS institutes for their contributions to the success of the CMS effort. In addition, we gratefully acknowledge the computing centres and personnel of the Worldwide LHC Computing Grid and other centres for delivering so effectively the computing infrastructure essential to our analyses. Finally, we acknowledge the enduring support for the construction and operation of the LHC, the CMS detector, and the supporting computing infrastructure provided by the following funding agencies: SC (Armenia), BMBWF and FWF (Austria); FNRS and FWO (Belgium); CNPq, CAPES, FAPERJ, FAPERGS, and FAPESP (Brazil); MES and BNSF (Bulgaria); CERN; CAS, MoST, and NSFC (China); MINCIENCIAS (Colombia); MSES and CSF (Croatia); RIF (Cyprus); SENESCYT (Ecuador); MoER, ERC PUT and ERDF (Estonia); Academy of Finland, MEC, and HIP (Finland); CEA and CNRS/IN2P3 (France); BMBF, DFG, and HGF (Germany); GSRI (Greece); NKFIH (Hungary); DAE and DST (India); IPM (Iran); SFI (Ireland); INFN (Italy); MSIP and NRF (Republic of Korea); MES (Latvia); LAS (Lithuania); MOE and UM (Malaysia); BUAP, CINVESTAV, CONACYT, LNS, SEP, and UASLP-FAI (Mexico); MOS (Montenegro); MBIE (New Zealand); PAEC (Pakistan); MES and NSC (Poland); FCT (Portugal); MESTD (Serbia); MCIN/AEI and PCTI (Spain); MOSTR (Sri Lanka); Swiss Funding Agencies (Switzerland); MST (Taipei); MHESI and NSTDA (Thailand); TUBITAK and TENMAK (Turkey); NASU (Ukraine); STFC (United Kingdom); DOE and NSF (USA).

\hyphenation{Rachada-pisek} Individuals have received support from the Marie-Curie programme and the European Research Council and Horizon 2020 Grant, contract Nos.\ 675440, 724704, 752730, 758316, 765710, 824093, 884104, and COST Action CA16108 (European Union); the Leventis Foundation; the Alfred P.\ Sloan Foundation; the Alexander von Humboldt Foundation; the Science Committee, project no. 22rl-037 (Armenia); the Belgian Federal Science Policy Office; the Fonds pour la Formation \`a la Recherche dans l'Industrie et dans l'Agriculture (FRIA-Belgium); the Agentschap voor Innovatie door Wetenschap en Technologie (IWT-Belgium); the F.R.S.-FNRS and FWO (Belgium) under the ``Excellence of Science -- EOS" -- be.h project n.\ 30820817; the Beijing Municipal Science \& Technology Commission, No. Z191100007219010; the Ministry of Education, Youth and Sports (MEYS) of the Czech Republic; the Shota Rustaveli National Science Foundation, grant FR-22-985 (Georgia); the Deutsche Forschungsgemeinschaft (DFG), under Germany's Excellence Strategy -- EXC 2121 ``Quantum Universe" -- 390833306, and under project number 400140256 - GRK2497; the Hellenic Foundation for Research and Innovation (HFRI), Project Number 2288 (Greece); the Hungarian Academy of Sciences, the New National Excellence Program - \'UNKP, the NKFIH research grants K 124845, K 124850, K 128713, K 128786, K 129058, K 131991, K 133046, K 138136, K 143460, K 143477, 2020-2.2.1-ED-2021-00181, and TKP2021-NKTA-64 (Hungary); the Council of Science and Industrial Research, India; the Latvian Council of Science; the Ministry of Education and Science, project no. 2022/WK/14, and the National Science Center, contracts Opus 2021/41/B/ST2/01369 and 2021/43/B/ST2/01552 (Poland); the Funda\c{c}\~ao para a Ci\^encia e a Tecnologia, grant CEECIND/01334/2018 (Portugal); the National Priorities Research Program by Qatar National Research Fund; MCIN/AEI/10.13039/501100011033, ERDF ``a way of making Europe", and the Programa Estatal de Fomento de la Investigaci{\'o}n Cient{\'i}fica y T{\'e}cnica de Excelencia Mar\'{\i}a de Maeztu, grant MDM-2017-0765 and Programa Severo Ochoa del Principado de Asturias (Spain); the Chulalongkorn Academic into Its 2nd Century Project Advancement Project, and the National Science, Research and Innovation Fund via the Program Management Unit for Human Resources \& Institutional Development, Research and Innovation, grant B05F650021 (Thailand); the Kavli Foundation; the Nvidia Corporation; the SuperMicro Corporation; the Welch Foundation, contract C-1845; and the Weston Havens Foundation (USA).
\end{acknowledgments}

\bibliography{auto_generated}

\clearpage
\appendix
\numberwithin{figure}{section}
\numberwithin{table}{section}
\section{Predicted and observed events} \label{app:table}

In this appendix, we present the numerical values and uncertainties for each of the signal and low-\ptmiss regions defined in Fig.~\ref{fig:Binning}.
These values correspond to the results presented in Fig.~\ref{fig:TotPredBkg}.

\begin{table}[h!]
\centering
\topcaption{The number of events predicted and observed for the signal regions and the low-\ptmiss regions used for the estimation of the \gjets and QCD multijet backgrounds.}
\renewcommand{\arraystretch}{1.2}
\cmsTable{
\label{table:pred}
\begin{tabular}{rBBBBBr}
\hline
\multicolumn{1}{c}{\ptmiss region [{\GeVns}]} & \multicolumn{4}{c}{Lost lepton} & \multicolumn{4}{c}{\Pe misidentified as \PGg} & \multicolumn{4}{c}{\zgamma} & \multicolumn{4}{c}{Multijet$+\PGg$} & \multicolumn{4}{c}{Total pred.} & \multicolumn{1}{c}{Observed}\\
\hline
\multicolumn{22}{c}{$2 \leq \nj \leq 4$, $\nb = 0$}\\
\hline
200--300 & 731& & 64& & 459& & 97& & 730& & 146& & 946& & 267& & 2870& & 317& & 2865\\
300--370 & 174& & 19& & 116& & 26& & 265& & 59& & 71&.5 & 25&.8 & 626& & 72& & 641\\
370--450 & 76&.5 & 9&.7 & 58&.2 & 13&.0 & 143& & 36& & 25&.1 & 9&.0 & 303& & 40& & 325\\
450--600 & 33&.1 & 5&.5 & 36&.5 & 7&.7 & 105& & 34& & 12&.0 & 4&.3 & 186& & 36& & 157\\
600--750 & 14&.8 & 3&.6 & 10&.4 & 2&.3 & 21&.3 & 7&.3 & 1&.49 & 0&.54 & 48&.0 & 8&.8 & 32\\
750--900 & 4&.17 & 1&.65 & 3&.23 & 0&.72 & 11&.8 & 6&.4 & \multicolumn{1}{r@{\hspace{0em}}}{0} & \multicolumn{1}{l@{$\,\vphantom{0}^{+}_{-}\,$}}{} & \multicolumn{1}{r@{\hspace{0em}}}{$\vphantom{0}^{0}_{0}$} & \multicolumn{1}{@{\hspace{0em}}l}{$\vphantom{0}^{.00360}$} & 19&.2 & 6&.4 & 19\\
${\geq}900$ & 2&.59 & 1&.33 & 1&.71 & 0&.50 & 3&.00 & 1&.73 & 0&.491 & 0&.177 & 7&.79 & 2&.16 & 11\\
\hline
\multicolumn{22}{c}{$5 \leq \nj \leq 6$, $\nb = 0$}\\
\hline
200--300 & 68&.3 & 8&.0 & 42&.7 & 9&.5 & 8&.96 & 2&.62 & 88&.6 & 7&.8 & 209& & 14& & 208\\
300--370 & 15&.8 & 3&.4 & 14&.5 & 3&.3 & 4&.60 & 1&.71 & 4&.17 & 0&.88 & 39&.0 & 4&.7 & 41\\
370--450 & 7&.00 & 2&.24 & 7&.39 & 1&.63 & 4&.49 & 1&.85 & 3&.79 & 0&.80 & 22&.7 & 3&.3 & 21\\
450--600 & 7&.00 & 2&.10 & 5&.84 & 1&.44 & 2&.90 & 1&.77 & 1&.92 & 0&.41 & 17&.7 & 3&.1 & 22\\
600--750 & 1&.75 & 1&.10 & 2&.15 & 0&.54 & 0&.831 & 1&.009 & 0&.264 & 0&.056 & 5&.00 & 1&.61 & 4\\
${\geq}750$ & 2&.33 & 1&.22 & 0&.779 & 0&.254 & 0&.754 & 1&.014 & 1&.01 & 0&.21 & 4&.87 & 1&.61 & 0\\
\hline
\multicolumn{22}{c}{$\nj \geq 7$, $\nb = 0$}\\
\hline
200--300 & 7&.44 & 2&.01 & 8&.20 & 1&.16 & 1&.77 & 0&.87 & 12&.1 & 1&.9 & 29&.5 & 3&.2 & 29\\
300--370 & 4&.19 & 1&.59 & 2&.28 & 0&.35 & 0&.717 & 0&.580 & 0&.00729 & 0&.00370 & 7&.19 & 1&.70 & 5\\
370--450 & 1&.40 & 0&.88 & 1&.72 & 0&.29 & 0&.200 & 0&.124 & 0&.362 & 0&.184 & 3&.68 & 0&.97 & 1\\
450--600 & 1&.40 & 0&.79 & 1&.22 & 0&.27 & 0&.0115 & 0&.0139 & 0&.515 & 0&.261 & 3&.14 & 0&.86 & 2\\
${\geq}600$ & 0&.931 & 0&.786 & 0&.695 & 0&.161 & 0&.0271 & 0&.0397 & \multicolumn{1}{r@{\hspace{0em}}}{0} & \multicolumn{1}{l@{$\,\vphantom{0}^{+}_{-}\,$}}{} & \multicolumn{1}{r@{\hspace{0em}}}{$\vphantom{0}^{0}_{0}$} & \multicolumn{1}{@{\hspace{0em}}l}{$\vphantom{0}^{.00508}$} & 1&.66 & 0&.81 & 1\\
\hline
\multicolumn{22}{c}{$2 \leq \nj \leq 4$, $\nb \geq 1$}\\
\hline
200--300 & 238& & 22& & 139& & 29& & 57&.0 & 19&.6 & 241& & 91& & 675& & 94& & 674\\
300--370 & 53&.9 & 7&.9 & 30&.3 & 6&.3 & 19&.1 & 7&.6 & 14&.4 & 6&.6 & 118& & 14& & 114\\
370--450 & 19&.1 & 4&.0 & 14&.0 & 3&.0 & 8&.35 & 3&.90 & 4&.52 & 2&.08 & 46&.0 & 6&.4 & 58\\
450--600 & 13&.0 & 3&.4 & 7&.45 & 1&.64 & 7&.80 & 4&.08 & 1&.86 & 0&.85 & 30&.1 & 5&.5 & 35\\
${\geq}600$ & 3&.47 & 1&.56 & 2&.30 & 0&.58 & 3&.00 & 2&.31 & 0&.242 & 0&.111 & 9&.02 & 2&.73 & 6\\
\hline
\multicolumn{22}{c}{$5 \leq \nj \leq 6$, $\nb \geq 1$}\\
\hline
200--300 & 77&.6 & 9&.8 & 62&.4 & 13&.6 & 1&.55 & 1&.09 & 53&.2 & 10&.0 & 195& & 19& & 194\\
300--370 & 17&.2 & 3&.9 & 16&.7 & 3&.7 & 1&.26 & 0&.77 & 7&.57 & 2&.33 & 42&.7 & 5&.9 & 48\\
370--450 & 8&.24 & 2&.46 & 7&.31 & 1&.59 & 0&.672 & 0&.633 & 1&.54 & 0&.48 & 17&.8 & 3&.1 & 23\\
450--600 & 2&.06 & 1&.11 & 4&.25 & 0&.95 & 0&.0772 & 0&.0616 & \multicolumn{1}{r@{\hspace{0em}}}{0} & \multicolumn{1}{l@{$\,\vphantom{0}^{+}_{-}\,$}}{} & \multicolumn{1}{r@{\hspace{0em}}}{$\vphantom{0}^{0}_{0}$} & \multicolumn{1}{@{\hspace{0em}}l}{$\vphantom{0}^{.00308}$} & 6&.39 & 1&.46 & 8\\
${\geq}600$ & 2&.06 & 1&.15 & 1&.27 & 0&.29 & 0&.0587 & 0&.0452 & 1&.41 & 0&.44 & 4&.81 & 1&.22 & 3\\
\hline
\multicolumn{22}{c}{$\nj \geq 7$, $\nb \geq 1$}\\
\hline
200--300 & 18&.3 & 4&.0 & 24&.2 & 5&.1 & 0&.0767 & 0&.0579 & 22&.1 & 7&.5 & 64&.6 & 10&.0 & 64\\
300--370 & 5&.89 & 2&.02 & 7&.14 & 1&.65 & 0&.697 & 0&.567 & 1&.48 & 1&.02 & 15&.2 & 2&.9 & 8\\
370--450 & 4&.12 & 1&.58 & 3&.31 & 0&.71 & 0&.0600 & 0&.0555 & 0&.573 & 0&.395 & 8&.07 & 1&.76 & 9\\
450--600 & 2&.95 & 1&.36 & 2&.04 & 0&.47 & \multicolumn{1}{r@{\hspace{0em}}}{0} & \multicolumn{1}{l@{$\,\vphantom{0}^{+}_{-}\,$}}{} & \multicolumn{1}{r@{\hspace{0em}}}{$\vphantom{0}^{0}_{0}$} & \multicolumn{1}{@{\hspace{0em}}l}{$\vphantom{0}^{.00472}$} & 0&.364 & 0&.251 & 5&.36 & 1&.48 & 3\\
${\geq}600$ & 1&.18 & 0&.84 & 0&.581 & 0&.223 & 0&.0270 & 0&.0283 & \multicolumn{1}{r@{\hspace{0em}}}{0} & \multicolumn{1}{l@{$\,\vphantom{0}^{+}_{-}\,$}}{} & \multicolumn{1}{r@{\hspace{0em}}}{$\vphantom{0}^{0}_{0}$} & \multicolumn{1}{@{\hspace{0em}}l}{$\vphantom{0}^{.00689}$} & 1&.80 & 0&.83 & 1\\
\hline
\multicolumn{22}{c}{\vtag}\\
\hline
200--300 & 172& & 17& & 174& & 35& & 39&.2 & 8&.1 & 180& & 51& & 565& & 63& & 564\\
300--370 & 47&.8 & 8&.4 & 34&.9 & 7&.6 & 11&.6 & 3&.6 & 8&.57 & 4&.06 & 103& & 13& & 97\\
370--450 & 19&.8 & 4&.9 & 13&.0 & 2&.9 & 9&.80 & 3&.38 & 3&.60 & 1&.71 & 46&.2 & 7&.2 & 52\\
450--600 & 12&.5 & 3&.2 & 6&.02 & 1&.37 & 8&.48 & 3&.55 & 0&.952 & 0&.451 & 27&.9 & 5&.2 & 36\\
600--750 & 7&.28 & 2&.91 & 1&.38 & 0&.34 & 2&.88 & 1&.72 & 0&.334 & 0&.158 & 11&.9 & 3&.4 & 4\\
${\geq}750$ & 2&.08 & 1&.37 & 0&.670 & 0&.178 & 1&.78 & 1&.52 & \multicolumn{1}{r@{\hspace{0em}}}{0} & \multicolumn{1}{l@{$\,\vphantom{0}^{+}_{-}\,$}}{} & \multicolumn{1}{r@{\hspace{0em}}}{$\vphantom{0}^{0}_{0}$} & \multicolumn{1}{@{\hspace{0em}}l}{$\vphantom{0}^{.00473}$} & 4&.54 & 2&.01 & 2\\
\hline
\multicolumn{22}{c}{\htag}\\
\hline
200--300 & 76&.3 & 10&.3 & 88&.2 & 18&.7 & 14&.1 & 3&.7 & 67&.5 & 15&.3 & 246& & 27& & 245\\
300--370 & 24&.1 & 5&.4 & 24&.2 & 5&.1 & 5&.73 & 2&.20 & 6&.75 & 2&.94 & 60&.7 & 8&.3 & 60\\
370--450 & 7&.02 & 2&.83 & 10&.5 & 2&.2 & 5&.28 & 2&.41 & 2&.83 & 1&.23 & 25&.6 & 4&.5 & 34\\
450--600 & 9&.03 & 2&.98 & 4&.62 & 1&.05 & 4&.02 & 2&.15 & \multicolumn{1}{r@{\hspace{0em}}}{0} & \multicolumn{1}{l@{$\,\vphantom{0}^{+}_{-}\,$}}{} & \multicolumn{1}{r@{\hspace{0em}}}{$\vphantom{0}^{0}_{0}$} & \multicolumn{1}{@{\hspace{0em}}l}{$\vphantom{0}^{.00436}$} & 17&.7 & 3&.8 & 20\\
600--750 & 3&.01 & 1&.76 & 1&.02 & 0&.28 & 2&.56 & 1&.34 & 0&.706 & 0&.308 & 7&.30 & 2&.28 & 2\\
${\geq}750$ & 2&.01 & 1&.40 & 0&.504 & 0&.144 & 1&.19 & 1&.05 & \multicolumn{1}{r@{\hspace{0em}}}{0} & \multicolumn{1}{l@{$\,\vphantom{0}^{+}_{-}\,$}}{} & \multicolumn{1}{r@{\hspace{0em}}}{$\vphantom{0}^{0}_{0}$} & \multicolumn{1}{@{\hspace{0em}}l}{$\vphantom{0}^{.00436}$} & 3&.72 & 1&.66 & 2\\
\hline
\end{tabular}
}
\end{table}
\cleardoublepage \section{The CMS Collaboration \label{app:collab}}\begin{sloppypar}\hyphenpenalty=5000\widowpenalty=500\clubpenalty=5000\input{SUS-21-009-public-authorlist.tex}\end{sloppypar}
\end{document}

%% file: SUS-21-009-public-authorlist.tex
\cmsinstitute{Yerevan Physics Institute, Yerevan, Armenia}
{\tolerance=6000
A.~Hayrapetyan, A.~Tumasyan\cmsAuthorMark{1}\cmsorcid{0009-0000-0684-6742}
\par}
\cmsinstitute{Institut f\"{u}r Hochenergiephysik, Vienna, Austria}
{\tolerance=6000
W.~Adam\cmsorcid{0000-0001-9099-4341}, J.W.~Andrejkovic, T.~Bergauer\cmsorcid{0000-0002-5786-0293}, S.~Chatterjee\cmsorcid{0000-0003-2660-0349}, K.~Damanakis\cmsorcid{0000-0001-5389-2872}, M.~Dragicevic\cmsorcid{0000-0003-1967-6783}, A.~Escalante~Del~Valle\cmsorcid{0000-0002-9702-6359}, P.S.~Hussain\cmsorcid{0000-0002-4825-5278}, M.~Jeitler\cmsAuthorMark{2}\cmsorcid{0000-0002-5141-9560}, N.~Krammer\cmsorcid{0000-0002-0548-0985}, D.~Liko\cmsorcid{0000-0002-3380-473X}, I.~Mikulec\cmsorcid{0000-0003-0385-2746}, J.~Schieck\cmsAuthorMark{2}\cmsorcid{0000-0002-1058-8093}, R.~Sch\"{o}fbeck\cmsorcid{0000-0002-2332-8784}, D.~Schwarz\cmsorcid{0000-0002-3821-7331}, M.~Sonawane\cmsorcid{0000-0003-0510-7010}, S.~Templ\cmsorcid{0000-0003-3137-5692}, W.~Waltenberger\cmsorcid{0000-0002-6215-7228}, C.-E.~Wulz\cmsAuthorMark{2}\cmsorcid{0000-0001-9226-5812}
% ...
\par}
\vskip\cmsinstskip
\dag:~Deceased\\
$^{1}$Also at Yerevan State University, Yerevan, Armenia\\
$^{2}$Also at TU Wien, Vienna, Austria\\
% ...
